\documentclass[aps,prb,twocolumn,floatfix,superscriptaddress,longbibliography,nofootinbib,10pt]{revtex4-2}

\usepackage{graphicx, color, soul}
\usepackage{amsmath, amsfonts, amssymb, mathrsfs, dsfont, mathtools}

\usepackage[dvipsnames, pdftex, table]{xcolor}
\usepackage[normalem]{ulem} 
\usepackage[mathscr]{eucal}
\usepackage{bm}
\usepackage{tabularx, booktabs}
\usepackage{multirow}
\usepackage{physics}
\usepackage{bbold}
\usepackage{comment}
\usepackage{wasysym}
\usepackage{tikz}    
\usepackage{stackengine}
\usepackage{csquotes}
\usepackage[colorlinks, breaklinks, 
            linkcolor=Blue,
            citecolor=Blue,
            urlcolor=Blue]{hyperref}
            
\usepackage[all]{hypcap} 
\usepackage{enumitem}
\usepackage{placeins}

\stackMath

\newcommand{\eq}[1]{\begin{align}#1\end{align}}

\date{\today}

\begin{document}
\title{Exactly solvable spin liquids in Kitaev bilayers and moir\'e superlattices}

\author{Ivan Dutta}
\affiliation{National Institute of Science Education and Research, Jatni, 752050, India}
\affiliation{Homi Bhabha National Institute, Training School Complex, Anushakti Nagar, Mumbai 400094, India}
\author{Anamitra Mukherjee}
\affiliation{National Institute of Science Education and Research, Jatni, 752050, India}
\affiliation{Homi Bhabha National Institute, Training School Complex, Anushakti Nagar, Mumbai 400094, India}
\author{Onur Erten}
\affiliation{Department of Physics, Arizona State University, Tempe, Arizona 85287, USA}
\author{Kush Saha}
\affiliation{National Institute of Science Education and Research, Jatni, 752050, India}
\affiliation{Homi Bhabha National Institute, Training School Complex, Anushakti Nagar, Mumbai 400094, India}

\begin{abstract}
Building on the recent advancements on moir\'e superlattices, we propose an exactly solvable model with Kitaev-type interactions on a bilayer honeycomb lattice for both AA stacking and moir\'e superlattices.
Using Monte Carlo simulations and variational analysis, we uncover a rich variety of phases where the intra and interlayer $\mathbb{Z}_2$ fluxes (visons) are arranged in a periodic fashion in the ground state, tuned by interlayer coupling and out-of-plane external magnetic field.
We further extend our model to moir\'e superlattices at various commensurate twist angles around two distinct twist centers represented by $C_{3z}$ and $C_{6z}$ of the honeycomb lattice. 
Our simulations reveal generalized arrangements of plaquette values that correlate with the AA or AB stacking regions across the moir\'e unit cell. 
Moreover, we find that, depending on the twist angle, twist center and interlayer coupling, moir\'e superlattices exhibit to a variety of gapped and gapless spin liquid phases and can also host corner and edge modes. Our results highlight the rich physics in bilayer and twisted bilayer models of exactly solvable quantum spin liquids.

\end{abstract}
\maketitle


\section{Introduction}
Since Anderson's pioneering work on resonating valence bond states~\cite{Anderson_RVB}, followed by Kitaev's seminal paper~\cite{Kitaev_Model} on exactly solvable spin-1/2 honeycomb model, the field of quantum spin liquids (QSLs) \cite{Senthil_Science, Balents2010_Nature, Savary_2017, Moessner_Moore_2021, Mandal_Kitaev} continues to emerge as an active and promising area of research for both experimentalists and theorists. This is because QSLs are known to host a plethora of exotic properties such as long-range entanglement, topological order and fractionalization of elementary excitations~\cite{Zhou_RMP, Roderich_Annual_Review, Wen_2017, hermanns2018}. Further, the exactly solvable QSL models provide deep insights into several essential physical properties such as dynamical spin correlations, transport phenomena and thermodynamic properties~\cite{Knolle_PRL_2014, Baskaran_2007, Nasu_PRB_2015, Nasu_PRL_2017, Alexandros_PRB_2017} as already demonstrated in various two and three dimensional models~\cite{Kitaev_Model, Yang_2007, Yao_2007_PRL, Yao_PRL_2009, Saptarshi_2009, baskaran2009exact, Wu_Gamma_PRB, Tikhonov_PRL_2010, Chua_PRB_2011, Furusaki_PRB_2012, Chulliparambil_2020, Akram_Vison_2023, Keskiner_PRB_2023}.  Despite numerous proposals on candidate materials such as iridates ~\cite{hwan2015_nature, Kitagawa2018_nature}, $\alpha$-RuCl$_3$~\cite{Takagi2019_Rucl3} and van der Waals (vdW) materials~\cite{Chris_2020_VdW, blei2021_AIP}, where Kitave-type interactions are predicted to be strong, the lack of natural materials exhibiting definitive QSL signatures keeps the search active and ongoing.   
 
While the study of monolayer QSL models has received tremendous attention since their initial predictions, recent years have witnessed a surge in studying bilayer QSLs~\cite{Nussinov_PRB_2009, Tomishige_PRB_2019, Taylor_Hughes_2020, Haskell_PRB_2022, Vijayvargia_PRR_2023, Nica2023_NPJ, Vijayvargia_PRB_2024,Vijayvargia_arXiv2025,Seifert_PRB_2018,toriccodebilayer_PRB,toriccodekilayer_PRB,samimi2024quantumphasediagrambilayer}. This renewed focus is largely inspired by the studies on moir\'e superlattices of graphene and other two-dimensional materials, which have attracted considerable attention due to a series of unconventional correlated phenomena, including magic-angle flat-bands and unconventional superconductivity~\cite{TBG_Suarez_2010, TBG_Mele_2010, Andrei2020_review, Macdonald_TBG_PNAS, Cao_2018_Nature}. In these twisted systems, the resulting moir\'e superlattice structure offers a highly tunable platform, where lattice mismatches and twist angles act as versatile control parameters. This flexibility enables the engineering of exotic quantum states and emergent phenomena that may not be achievable in conventional models. In view of this, bilayer QSLs present a promising direction to explore exact solvability of various lattice models, their ground state properties in the field free and finite field cases. This in turn may provide insights into the interplay of strong correlations, topology, and quantum entanglement. 

\begin{figure}[!t]
    \centering
    \includegraphics[width=1\columnwidth]{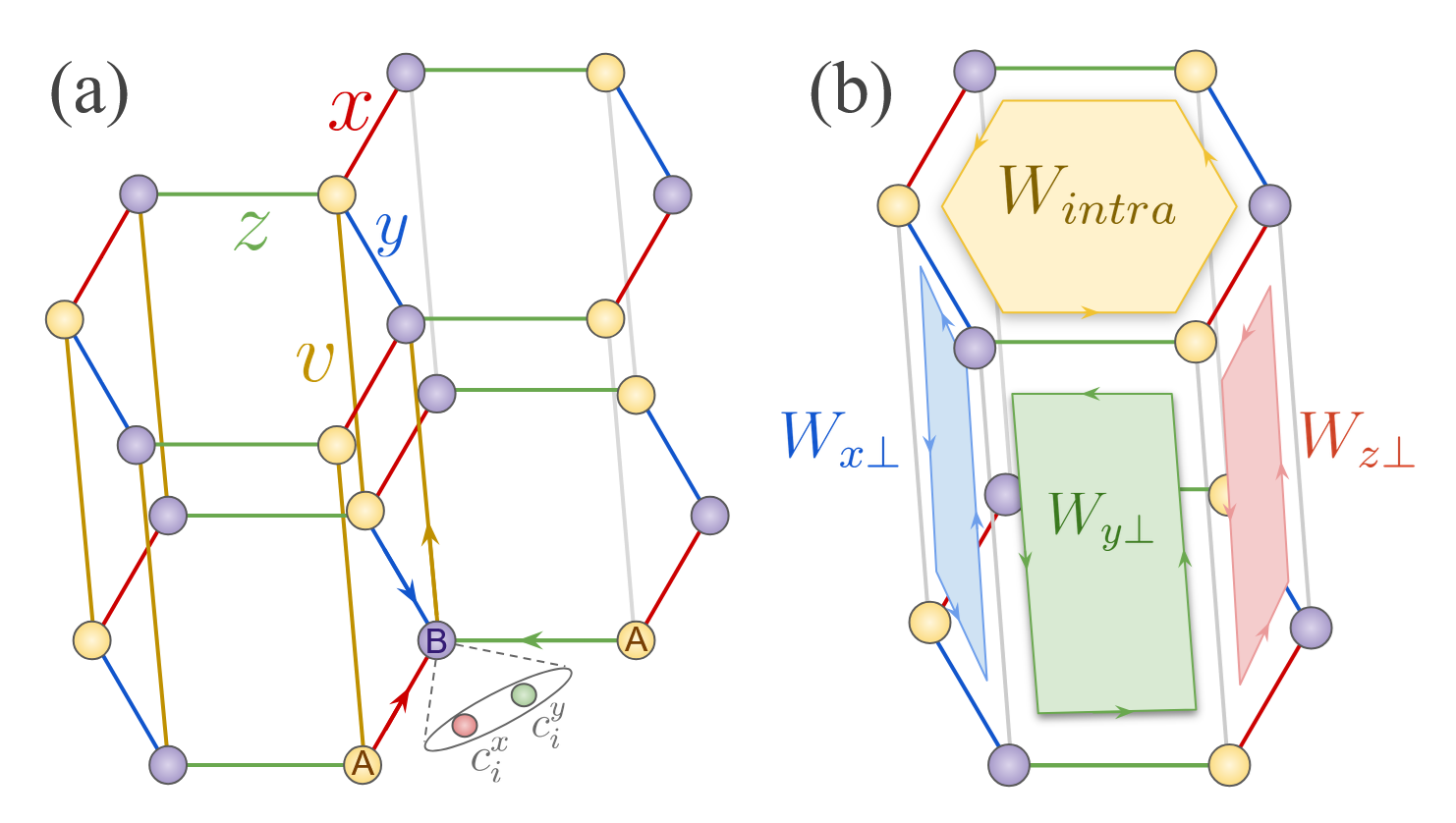}
        \caption{Schematics of the bilayer Kitaev-like lattice. (a) The lattice is composed of two sublattices A and B represented by yellow and violet spheres respectively. From each site, four distinct Kitaev-type bonds are depicted in red ($x$), blue ($y$), green ($z$), and yellow ($v$) colors. The $c_i^x$ and $c_i^y$ below represent two free Majorana fermions obtained by decomposing the spin Hamiltonian into fermionic counterpart. (b) The lattice features one intralayer hexagonal plaquette $W_{\text{intra}}$ and three distinct interlayer square plaquettes ($W_{x \perp}, W_{y \perp}, W_{z \perp}$).
        } 
    \label{fig:dis_surface}
\end{figure}
 
Here, we propose a bilayer Kitaev-like model by generalizing bond-dependent interactions in terms of $\Gamma$ matrices. Unlike previous studies on bilayer QSLs~\cite{Nussinov_PRB_2009,Seifert_PRB_2018, Tomishige_PRB_2019, Taylor_Hughes_2020, Haskell_PRB_2022, Vijayvargia_PRR_2023, Nica2023_NPJ, Vijayvargia_PRB_2024, Vijayvargia_arXiv2025, samimi2024quantumphasediagrambilayer}, our models remain exactly solvable even in the presence of an external out-of-plane magnetic field and a finite twist. Using variational analysis and Monte Carlo simulations~\cite{Yunoki_MC_1998,Churchill_MC_2025,Batista_MC_2019}, we first chart out flux configurations of the ground states as determined by the distinct fluxes through the intralayer hexagon plaquettes and interlayer square plaquettes of the bilayer for different values of out-of-plane magnetic field and interlayer coupling of the simple AA stacked bilayer model. We then extend this to moir\'e superlattices at different commensurate twist angles for distinct twist centers as defined by $C_{3z}$ (lattice site) and $C_{6z}$ (plaquette center) invariant points of the honeycomb lattice. Unlike the AA stacking pattern, the commensurate twist introduces both AA and AB stacking regions within the unit cell. Interestingly, every square plaquette in AA region acquires $\pi$ flux while those in AB region acquire $0$ flux. This reveals a generalized dependence of plaquette values on the local stacking configurations in the superlattice, regardless of twist angle and twist centers. Moreover, odd-sided interlayer plaquettes such as pentagons and heptagons also appear within the unit cell, which in turn spontaneously breaks time-reversal symmetry of the Majorana Hamiltonian. We further find that the ground states of the twisted model can lead to a variety of gapped and gapless phases with varying interlayer coupling and twist angles. Additionally, we show that the bilayer moir\`e superlattices may host floating boundary modes analogous to boundary modes reported recently in a few lattice models~\cite{FEB2025,FragilityAltland2024,Cook2022}. The appearance of these modes is attributed to broken time-reversal symmetry by odd-number of plaquettes. 
 
The rest of the paper is organized as follows. In Sec.~\ref{sec:AA_stacked}A, we introduce the model, discuss mutually commuting plaquette operators and the formalism. This is followed by Sec.~\ref{sec:AA_stacked}B where we discuss ground state of the AA-stacked bilayer spin liquids under both zero and finite out-of-plane magnetic field. In Sec.~\ref{sec:moire}A, we discuss construction of moir\'e superlattices, incorporating finite twist between the two layers about two distinct twist centers $C_{3z}$ and $C_{6z}$ of the honeycomb lattice. We then find the ground states for various twist angles and discuss the possible gapped and gapless spin liquid phases in Sec.~\ref{sec:moire}B and  Sec.~\ref{sec:moire}C. This is followed by discussions on edge modes for different twist angles in Sec.~\ref{sec:moire}D. Finally, we conclude with a summary and outlook in  Sec.~\ref{Section_4}.

\section{\label{Section_2} AA stacked bilayer spin liquids}\label{sec:AA_stacked}

\subsection{Model and methods}

We consider an exactly solvable spin model with Kitaev-like interactions on a bilayer honeycomb lattice with AA stacking. The Hamiltonian is characterized by bond-dependent interactions with three types of intralayer ($x$, $y$, $z$) and one type of interlayer bonds ($v$) as depicted in Fig.~\ref{fig:dis_surface}a. Labeling these four distinct bonds from 1 to 4 respectively, the Hamiltonian can be expressed in terms of $\Gamma$ matrices as $H = H_{\text{intra}} + H_{\text{inter}}$ where, 
\begin{align}
&H_{\text{intra}} = \sum_{\langle jk \rangle _{\gamma},\nu} K_\gamma\; (\Gamma^{\gamma}_{\nu j}\,\Gamma^{\gamma}_{\nu k}\; +\;\Gamma^{\gamma5}_{\nu j}\,\Gamma^{\gamma5}_{\nu k}),\\
&H_{\text{inter}} = J\;\sum_{j} (\Gamma^{4}_{1j}\,\Gamma^{4}_{2j}\; +\;\Gamma^{45}_{1j}\,\Gamma^{45}_{2j})
\label{eq:Gamma_Model}.
\end{align}
Here, $K_{\gamma}$ represents the nearest-neighbour (NN) coupling along $\gamma \in \{x, y, z\}$ bond in each layer and $J$ represents interlayer NN coupling constant. The first subscript $\nu \in \{1,2\}$ denotes the layer index and second indices $j,k$ refer to lattice sites. Physically, these matrices can be interpreted as higher-order spin multipole operators or as a combination of spin ($\sigma$) and orbital ($\tau$) degrees of freedom. For the latter, one possible representation is, $\Gamma^\gamma = - \sigma^y \otimes \tau^\gamma $, $\Gamma^4 =  \sigma^x \otimes \mathbb{I}_2$, $\Gamma^5 =  -\sigma^z \otimes \mathbb{I}_2$ and $\Gamma^{\gamma \gamma'} = \frac{i}{2}[\Gamma^{\gamma}, \Gamma^{\gamma'}]$, satisfying the Clifford algebra $\{\Gamma^{\gamma}, \Gamma^{\gamma'}\} = 2 \delta^{\gamma \gamma'}$. The Hamiltonian possesses a rich set of local integrals of motion, i.e., plaquette operators $W$ that substantially simplify the problem. We identify four sets of such mutually commuting plaquette operators: one for intralayer $W_{\nu, \text{intra}}$ and three for interlayer $W_{\gamma \perp}$ plaquettes, as shown in Fig.~\ref{fig:dis_surface}b. Explicitly, they can be expressed as
\begin{align}
&W_{\nu, \text{ intra}} = -\prod_{\langle jk \rangle_{\gamma} \in\, \hexagon}\; \Gamma^{\gamma}_{\nu j}\,\Gamma^{\gamma}_{\nu k}  \nonumber\\
&W_{\gamma \perp} =- \prod_{\nu, \nu',\langle jk \rangle_{\gamma} \in\, \square}\; \Gamma_{\nu j}^{\gamma}\,\Gamma_{\nu' k}^{\gamma},
\label{eq:plaquette_operator}
\end{align} 
where the product is taken in anticlockwise direction. The plaquette operators commute with $H$ and have eigenvalues $\pm 1$ which allow the total Hilbert space of the Hamiltonian to be decomposed as a direct sum of the eigenspaces of every possible combination of plaquette eigenvalues.

The Hamiltonian can be diagonalized by introducing $6$ Majorana fermions per site, $\Gamma_j^\alpha = i\,b_j^\alpha\,c_j$ ($\alpha = 1,2,3,4,5$). This spin to fermion decomposition leads to a system featuring two distinct flavors of free Majorana fermions, while the other Majoranas emerge as static background $\mathbb{Z}_2$ gauge field as illustrated in Appendix~\ref{Appendix_A}. By relabeling $b_j^5 \rightarrow c_j^x$ and $c_j \rightarrow c_j^y$, the total Hamiltonian can be rewritten as, 
 \eq{
\mathcal{H} = \mathcal{H}_{\text{intra}} + \mathcal{H}_{\text{inter}}
}
where
\begin{align}
&\mathcal{H}_{\text{intra}} =  -\, i\sum_{\nu, \langle jk\rangle \gamma} K_{\gamma}\;u_{\nu, jk}^{\gamma}\;(c^{x}_{\nu j}\,c^{x}_{\nu k}\; +\;c^{y}_{\nu j}\,c^{y}_{\nu k})\nonumber\\
&\mathcal{H}_{\text{inter}} = -\, iJ\;\sum_{j}\;u^{4}_{j}\;(c^{x}_{1j}\,c^{x}_{2j}\; +\;c^{y}_{1j}\,c^{y}_{2j})
\label{eq:Majorana_Gamma_Model}.
\end{align}
Here $u_{\nu, jk}^{\gamma}$ and $u^{4}_{j}$ are defined as intralayer and interlayer bond operators with eigenvalues $\pm 1$, hence acts as the $\mathbb{Z}_2$ gauge field. However, under this fermionization, the local Hilbert space becomes twice as large as the physical Hilbert space. To recover the physical subspace, we impose onsite parity constraint in each layer given by the condition $D_{\nu j} = \Gamma^1_{\nu j} \Gamma^2_{\nu j} \Gamma^3_{\nu j} \Gamma^4_{\nu j} \Gamma^5_{\nu j} = 1$. This eliminates unphysical states of the enlarged Hamiltonian via the projection operators $P_\nu = \Pi_i (1 + D_{\nu j})/2$ as $P\ket{\psi}_{\text{total}} = \ket{\psi_{\text{phys}}}$. 

Similar to the plaquette operators, the bond operators also mutually commute with each other and with the Hamiltonian. This enables to express the eigenstates of $\mathcal{H}$ as $\ket{\Psi} = \ket{\psi_{u}} \otimes \ket{u}$, where $\ket{u}$  refers to specific gauge field distributions of $u_{\nu, jk}^{\gamma}$ and $u^{4}_{j}$ and $\ket{\psi_{u}}$ is the corresponding eigenstate of $\mathcal{H}$ under that fixed u's. However, different configurations of u's may lead to same plaquette distributions. Locally, $\mathcal{H}$ is invariant under $\mathbb{Z}_2$ transformation in which $(c_j^x,c_j^y) \rightarrow - (c_j^x,c_j^y)$ and $u_{jk}^{\gamma} \rightarrow - u_{jk}^{\gamma}$. The $\mathbb{Z}_2$ gauge invariant Plaquette operators $W$ in Eq.~\ref{eq:plaquette_operator}, represented in terms of bond operators as
\eq{
W = \prod_{\langle jk \rangle \in \circlearrowleft} -i u_{jk},
\label{eq:Bond_operator}
}
identify all of the gauge equivalent eigenstates $\ket{\Psi}$ of $\mathcal{H}$ with the same emergent $\mathbb{Z}_2$ field as eigenvalues.

\begin{figure}[!t]
    \centering
    \includegraphics[width=0.98\columnwidth]{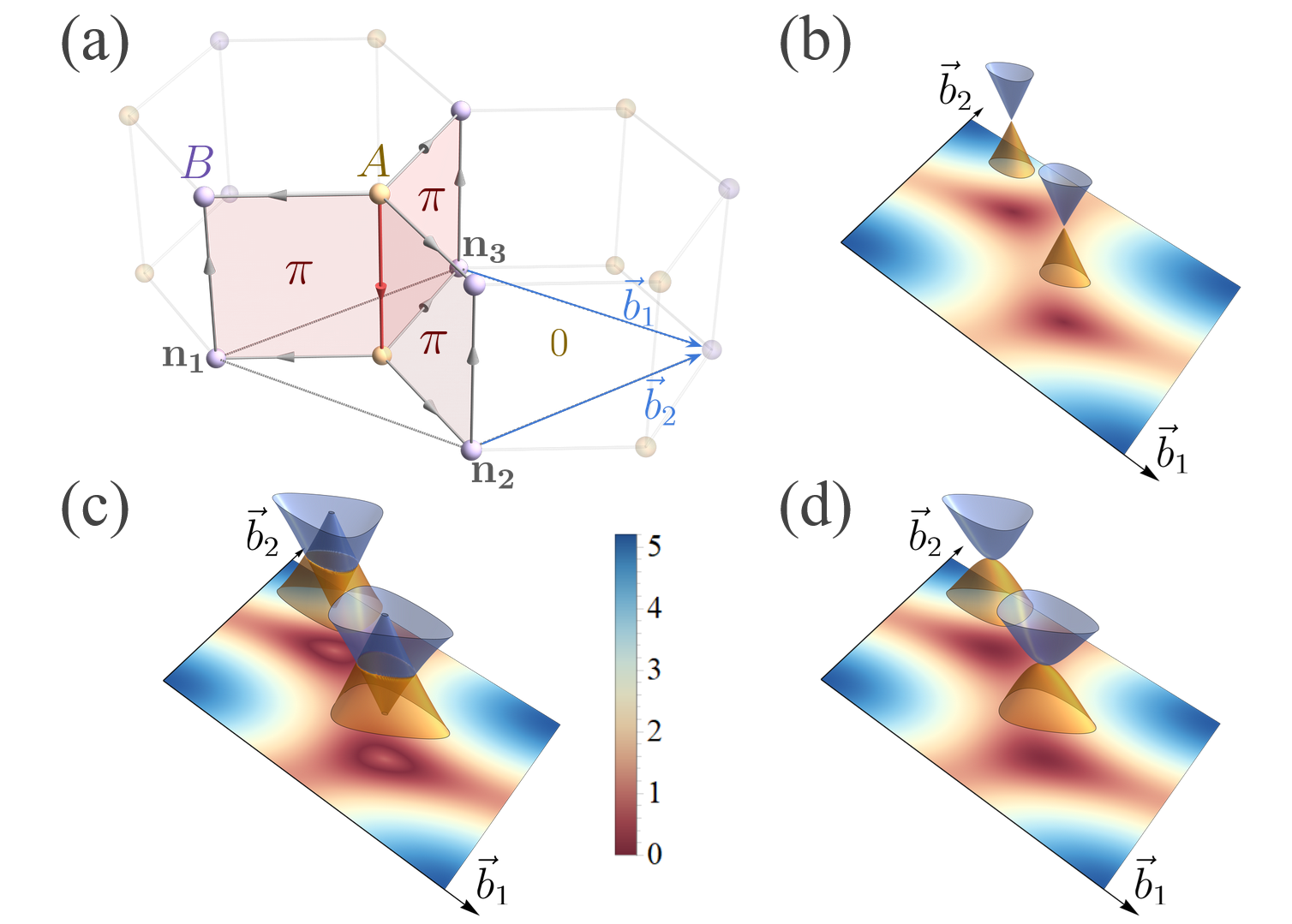}
        \caption{(a) The ground state $0$-$\pi$ flux configuration of the bilayer model with AA stacking. The arrows on the bonds indicate the direction corresponding to $u_{\langle ij \rangle} = +1$ where $\vec{b}_{1,2}$ are reciprocal lattice vectors. The energy spectrum (b) shows Dirac points for $K = 0$, $h_z = 0$, (c) nodal ring for $K = 0$, $h_z = 0.4$ and (d) quadratic band closing for $h_z = K = 0.4$. For $h_z \ne K > 0$ the spectrum remains gapped. In (b)-(d) the bands above show the dispersion of the middle two bands near zero energy while the colour plot below shows the full energy dispersion in the full BZ.
        } 
    \label{fig:Fig2}
\end{figure}
Next, we employ the Monte Carlo technique to determine the ground state flux phases of the model associated with the distribution of these eigenvalues. Our Monte Carlo approach involves the exact solution of Majorana Hamiltonian $\mathcal{H}$ in the background of static $\mathbb{Z}_2$ gauge fields. The free energy of the fermions, determined via diagonalization, is then used to thermally anneal the $\mathbb{Z}_2$ flux configurations using a standard  Metropolis algorithm. This approach has been successfully applied in earlier studies of vison crystals and the finite-temperature properties of the Kitaev model~\cite{Nasu_MC_2014, Batista_MC_2019, Feng_MC_2020, Ritwika_2024}. In our simulations, we consider a $16 \times 8$ lattice per layer, initializing the system at a high temperature with random flux configurations. We then gradually cooled the system, performing 3000 Monte Carlo sweeps at each temperature step. The ground-state flux configuration is estimated at a final temperature of $T = 10^{-3} K$.

\begin{figure}[!t]
    \centering
    \includegraphics[width=1.00\columnwidth]{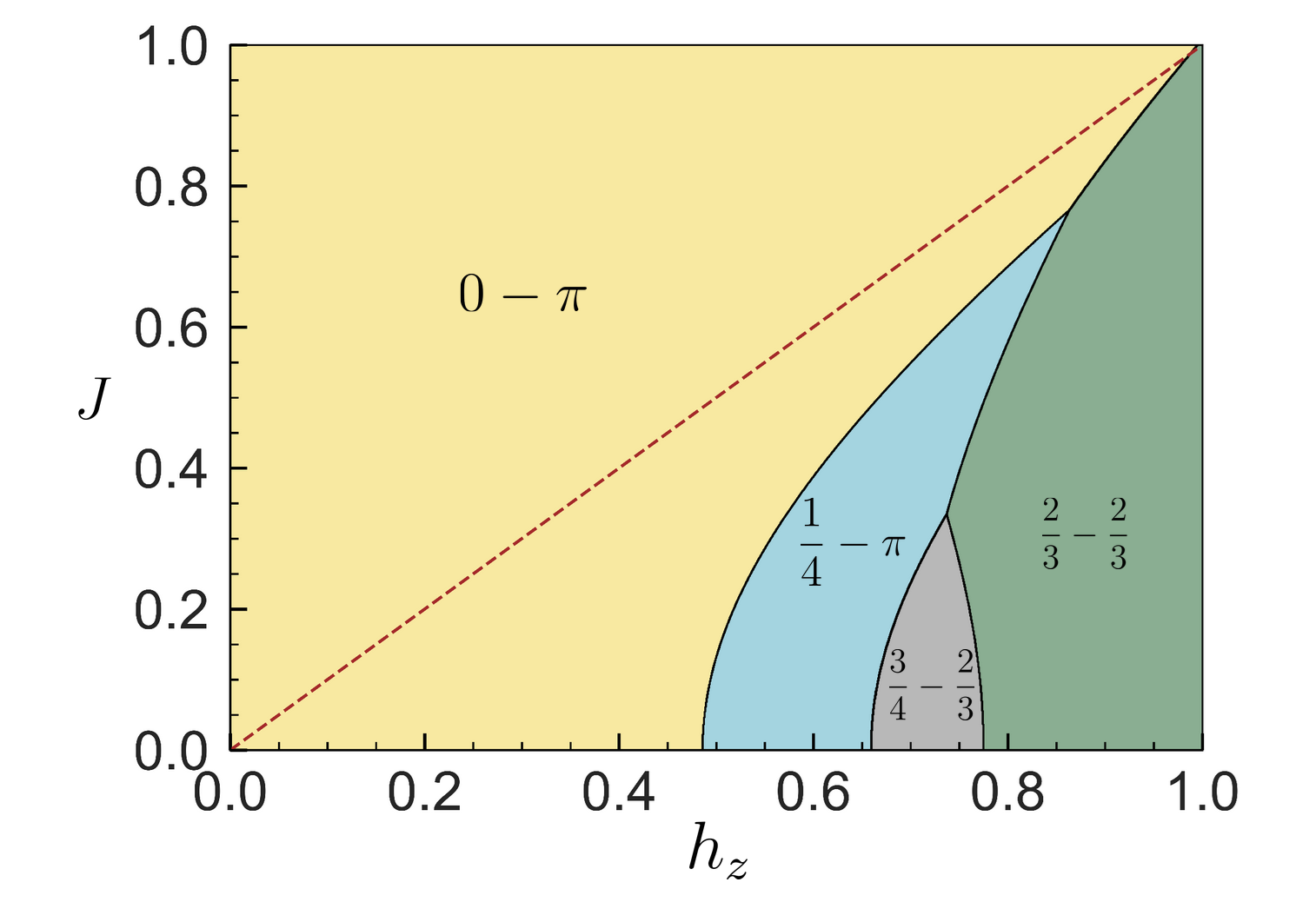}
        \caption{The $h_z/K$ vs. $J/K$ phase diagram of ground state configuration for the bilayer Hamiltonian. There exist four distinct phases separated by black colored phase boundary. These boundaries are evaluated variationally by calculating the energy of $192 \times 192$ system, for the flux configurations confirmed by Monte Carlo simulation. The brown dashed line in the $0-\pi$ phase shows the $h_z = J$ line, along which the band spectrum possesses quadratic band touching.
        } 
    \label{fig:phase}
\end{figure}

\subsection{Ground state phase diagram}

According the Lieb's theorem, the ground state of the monolayer model lies in zero-flux sector ($W_{\text{intra}} = 1$)~\cite{Lieb_1994}. As we introduce interlayer coupling, the Monte Carlo simulation predicts $\pi$-flux ($W_{\gamma \perp} = -1$) through all the interlayer plaquettes. This is in agreement with the previous findings~\cite{Chern_PRB_2010}. Thus the ground state in a bilayer setup is identified by both the intralayer and interlayer fluxes and we denote it for the zero field case as $0-\pi$, where $0$ and $\pi$ refer to intralayer and interlayer flux, respectively. 
\begin{figure*}[!t]
    \centering
    \includegraphics[width=0.99\textwidth]{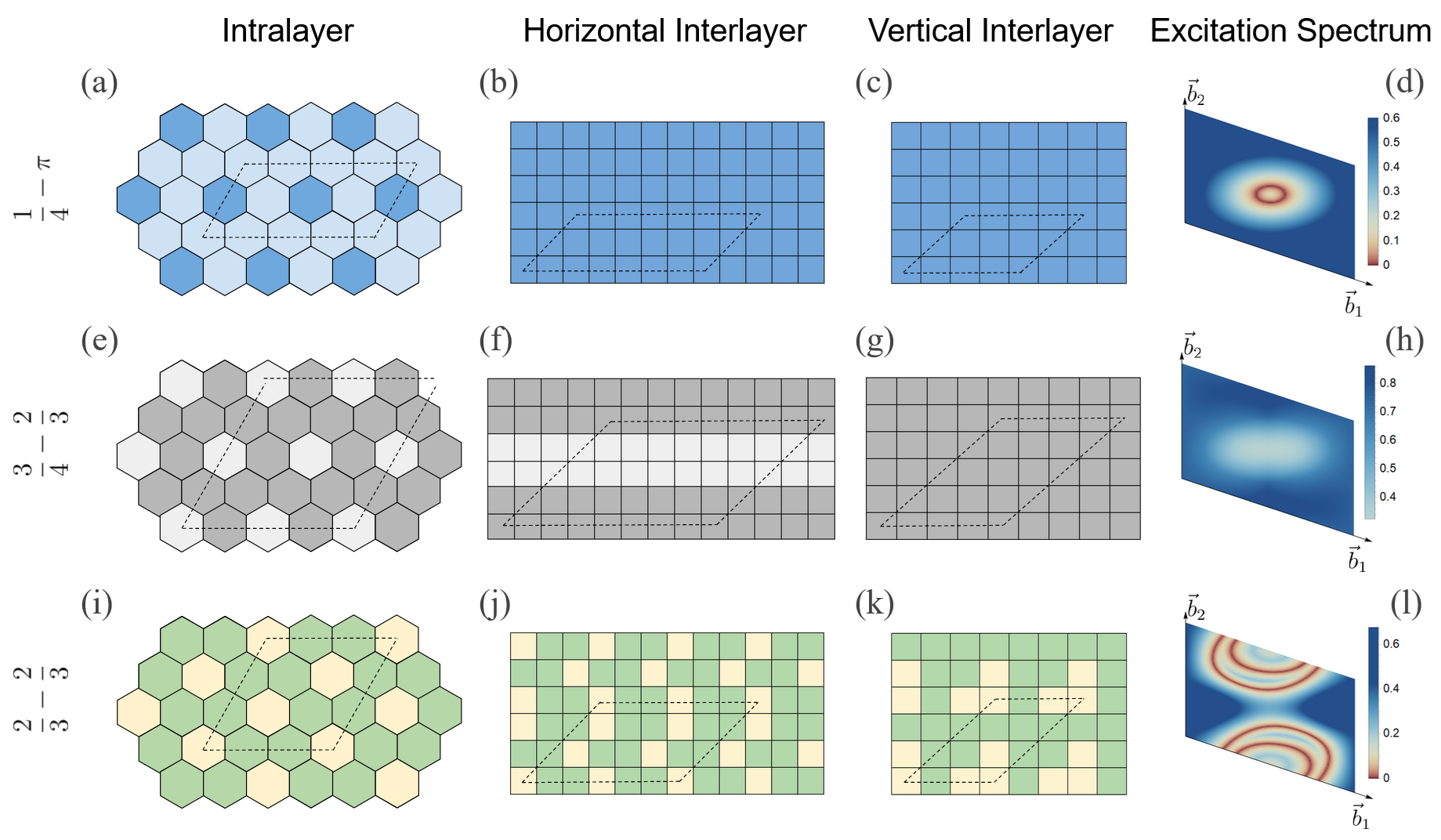}
        \caption{Four figures in each row show intralayer, horizontal interlayer, vertical interlayer flux distribution and corresponding excitation spectrum of the lowest conduction band respectively for (a-d) $1/4 - \pi$, (e-h) $3/4 - 2/3$ and (i-l) $2/3 - 2/3$ flux phases. A clear pictorial scheme to identify horizontal and vertical interlayer flux distribution is illustrated in Appendix~\ref{Appendix_A1}. We have chosen different colours of different flux phases for easy identification purposes in accordance with Fig.~\ref{fig:phase}. The unit cell of every flux phase is also marked by black dashed line. While the light coloured plaquettes in each figure of first three columns denote zero flux, the dark coloured plaquettes represent $\pi$ flux sectors in all the ground state configurations.}
    \label{fig:Different_plaquette}
\end{figure*}
The Majorana fermion excitation spectrum of this $0-\pi$ flux sector is obtained by fixing the gauge as $u^{\gamma}_{\nu, AB} = 1$ and interlayer bond operators as $u^{4}_{A(B)} = \mp 1$, where A, B refer to the sublattices as depicted in Fig.~\ref{fig:Fig2}a. Under this fixed gauge choice, the intralayer Hamiltonian in Eq.~\ref{eq:Majorana_Gamma_Model} for each Majorana fermion takes the form (assuming an isotropic intralayer coupling $K_{\gamma} = K$)
\eq{
\mathcal{H}_{\text{intra}}(\bm{k})  =\sum_{k \in \text{BZ}/2} \mathbb{I}_2 \, \otimes \{ \text{Re}f({\bm{k}})\, \Sigma_x + \text{Im}f(\bm{k})\, \Sigma_y \},
\label{eq:Intra_Model_0pi_Majorana}
}
where $f(\bm{k}) = -2iK\,(e^{i \bm{k}.\bm{n}_1}+e^{i \bm{k}.\bm{n}_2} + e^{i \bm{k}.\bm{n}_3})$ with lattice vectors $\bm{n}_{1,2} = (\pm \sqrt{3}/2, -1/2)a$, $\bm{n}_3 = (0,1)a$. $\Sigma$s are Pauli matrices representing the degrees of freedom of Majorana fermion. The sum of half Brillouin zone (BZ) is taken due to the Fourier transform properties of Majorana fermions. The identity matrix $\mathbb{I}_2$ accounts for the bilayer structure of the system. Similarly, the interlayer Hamiltonian in Eq. (\ref{eq:Majorana_Gamma_Model}) can be written as
\eq{
\mathcal{H}_{\text{inter}}(\bm{k})  = \sum_{k \in \text{BZ}/2} 2iJ\, \frac{\Sigma_x + i\Sigma_y}{2} \otimes  \, \Sigma_z + \text{H.c.}
\label{eq:Inter_Model_0pi_Majorana}
} 
The total Hamiltonian then takes the form $\mathcal{H}(\bm{k}) = \psi_{\bm{k}}^{\dag} \{(\mathcal{H}_{\text{intra}}(\bm{k}) + \mathcal{H}_{\text{inter}}(\bm{k})) \otimes \mathbb{1}_2\} \psi_{\bm{k}}$, where the identity matrix accounts for the presence of two distinct Majorana fermions. Here $\psi_{\bm{k}}$ is an eight component spinor defined as $\psi_{\bm{k}} = (c_{A,\bm{k},1}^x,c_{A,\bm{k},1}^y,c_{B,\bm{k},1}^x,...,c_{B,\bm{k},2}^y)^T$. The energy eigenvalues are  
\eq{
\epsilon_{1,2} = \pm \sqrt{|f(\bm{k})|^2 + 4J^2}
\label{eq:energy_zeropi}
}
with fourfold degeneracy. For $J = 0$ the dispersion spectrum exhibits  Dirac cones within the half BZ at $f(\bm{k}) = 0$, characteristic of the monolayer honeycomb model (Fig.~\ref{fig:Fig2}b).  However, finite $J$ induces a gap in the spectrum as $\epsilon_{1,2} \ne 0$.

To investigate the impact of an out-of-plane external magnetic field, we introduce an onsite Zeeman term,
\begin{align}
H_h = h_z \sum_{\nu,j} \Gamma_{\nu j}^5, 
\end{align}
where $h_z$ is the strength of external field. Notably, this term commutes with the Hamiltonian, preserving the exact solvability condition. In the Majorana representation, $H_h$ can be written as $\mathcal{H}_h = h_z \sum_{\nu,i} ic_{i \nu}^x c_{i \nu}^y$. This induces hybridization between the two Majorana fermions, which in turn breaks the degeneracy of the spectrum. In particular, the energy dispersion is obtained to be 
\eq{
\Tilde{\epsilon}_{1,2,3,4} = \pm (\epsilon_{1} \pm 2h_z)
\label{eq:energy_zeropi_magnetic}
}
with two-fold degeneracy. For $J=0$ the system hosts nodal ring-like gapless spectrum as illustrated in Fig.~\ref{fig:Fig2}c. Upon turning on a nonzero interlayer coupling $J$, the bulk spectrum gaps out for finite $h_z$, except $h_z = J$ line. Along this line, the spectrum shows quadratic band touching (Fig.~\ref{fig:Fig2}d).

For finite $\mathcal{H}_h$, Lieb's theorem does not hold. Accordingly, the ground state flux sector can shift from $0-\pi$ to a different pattern\cite{Batista_MC_2019, Chulliparambil_PRB2021, Ritwika_2024, Akram_Vison_2023}. To examine the stability of the $0-\pi$ phase as a function of $J$ and $h_z$, we perform Monte Carlo simulations and variational analysis. The phase diagram in Fig.~\ref{fig:phase} reveals that the $0-\pi$ flux phase is stable up to a critical $h_z$, above which new vison crystals emerge as the ground state of the system. We discover three new phases denoted by $1/4-\pi$, $3/4-2/3$ and $2/3-2/3$ in the $J-h_z$ plane as shown in Fig.~\ref{fig:phase}. While we use Monte Carlo simulations to discover these phases, the precise phase boundaries are determined by variational analysis.

The new flux phases in Fig.~\ref{fig:phase} are characterized by the ratio of number of $\pi$ flux plaquettes to the total number of plaquettes in a unit cell for both intra and interlayer plaquettes. Fig.~\ref{fig:Different_plaquette} illustrates flux patterns for these phases. The dark-colored plaquette refers to $\pi$ flux and light-colored plaquette refers to zero flux. Fig.~\ref{fig:Different_plaquette} (a,e,i) evidences that the three phases contain distinct numbers of finite $\pi$ flux intralayer plaquettes and the ratios discussed above are found to be $1/4$, $3/4$ and $2/3$, respectively. Contrary to the intralayer plaquettes, flux patterns for the interlayer vertical plaquettes are more challenging to represent graphically on 2D planes as they are oriented in different directions. This can be circumvented by dividing these interlayer plaquettes in two parts based on their attachments with the intralayer zigzag ($\bm{n}_{1,2}$) and vertical bonds ($\bm{n}_3$) of the honeycomb lattice as also indicated in Fig~\ref{fig:Fig2}a. A clear pictorial view of this scheme is also illustrated in Appendix~\ref{Appendix_A1}. Fig.~\ref{fig:Different_plaquette} (b,f,j) and (c,g,k) demonstrate flux patterns for both square plaquettes attached to both zigzag and vertical bonds of the intralayer lattice. Clearly, the total number of finite $\pi$ flux phases involving both these two types of vertical plaquettes differs for all three phases discussed before. In this case, the ratios are found to be $1$, $2/3$ and $2/3$, respectively. Note that the flux pattern further helps us to identify the unit cell and accordingly construct the Hamiltonian for each of these three phases in {\it momentum} space (see Appendix~\ref{Appendix_B}). The corresponding excitation spectra of the lowest conduction band are shown in Fig.~\ref{fig:Different_plaquette}. Clearly, both $1/4-\pi$ and $2/3-2/3$ exhibit gapless spectrum while $3/4-2/3$ is gapped. Both gapped phases $0-\pi$ and $3/4-2/3$, host abelian $Z_2$ topological order with zero Chern number. Note that we restrict our analysis to the parameter space $J, h_z \leq 1$, as beyond that limit the energy differences between flux configurations are relatively small to determine the correct ground state via Monte Carlo simulations. Additionally, the number of possible flux configurations is significantly larger in the bilayer model compared to the monolayer model to compute them variationally.

\section{\label{Section_3}moir\'e superlattices of spin liquids}\label{sec:moire}
\subsection{Model construction}

Next, we discuss the construction of our model for a moir\'e superlattice. We only consider commensurate twist angles given by~\cite{Shallcross_2010_TBGAngle,Cui_2024},
\eq{
\cos\theta_{p,q}  =\frac{3p^2 + 3pq + q^2/2}{3p^2 + 3pq + q^2}.
\label{eq:twistangle_formula}
}
Here $p$ and $q$ are coprime integers and $0 < \theta_{p,q} < \pi/3$. Unlike the AA stacking pattern, the local stacking changes within the moir\`e unit cell. The interlayer Hamiltonian in Eq. \ref{eq:Majorana_Gamma_Model} can be expressed as
\eq{
\mathcal{H}_{\text{inter}} = -\,iJ \;\sum_{\langle jj' \rangle} T_{jj'}(r) \; u^{4}_{j}\;(c^{x}_{1j}\,c^{x}_{2j'}\; +\;c^{y}_{1j}\,c^{y}_{2j'}),
}
where the hopping amplitude is defined as $T_{jj'}(r)$ between sites \( j \) and \( j' \) in adjacent layers with  bondlength $r \equiv \sqrt{(r^{\perp}_{jj'})^2 + (r^{l}_{jj'})^2}
$. Here, \( r^{\perp}_{jj'} = r^z_{j'} - r^z_{j} = a\) and the lateral distance \( r^{l}_{jj'} \) is computed from the lateral (x,y) coordinates of the two sites. We further introduce a cutoff \( r_{\rm c} \) such that 
$T_{jj'}(r) = e^{- r^{l}_{jj'}/a}$
, for \( r_{jj'} < r_{\rm c} \) following literature~\cite{Akram_2021_VdW, Akram_NanoLett2024, Akram_NanoLett2025}, and zero otherwise. The intralayer Hamiltonian \( \mathcal{H}_{\text{intra}} \) remains identical across all twist angles.

For vertically stacked sites as in the AA-stacked bilayers, \( r^{l}_{jj'} = 0 \). However, in moiré superlattices, the planar distances vary as a function of the local stacking configuration. Consequently, \( r^{l}_{jj'} \) takes a wide range of values, which we rank from shortest (strongest coupling) to longest (weakest coupling). With this we employ a scheme that ensures at most one interlayer bond per site controlled by the cutoff $r_{\rm c}$, as elaborated in Appendix~\ref{Appendix_C}. This construction in turn keeps the Hamiltonian exactly solvable even in the presence of $h_z$. Consequently, interlayer plaquettes with odd numbers of sides emerge around the sites with missing interlayer bonds in the unit cell and it breaks time-reversal symmetry as also highlighted in Ref.~\cite{Yao_2007_PRL,Wu_Gamma_PRB}. The definition of odd-sided plaquette operator is similar to even interlayer plaquettes in Eq.~\ref{eq:Bond_operator}, yielding $\pm i$ as eigenvalue. We note that while moir\'e superlattices of QSLs have been studied in recent years\cite{Taylor_Hughes_2020, Nica2023_NPJ, Haskell_PRB_2022}, our model stands out by being exactly solvable.

\begin{figure*}[!t]
    \centering
    \includegraphics[width=0.99\textwidth]{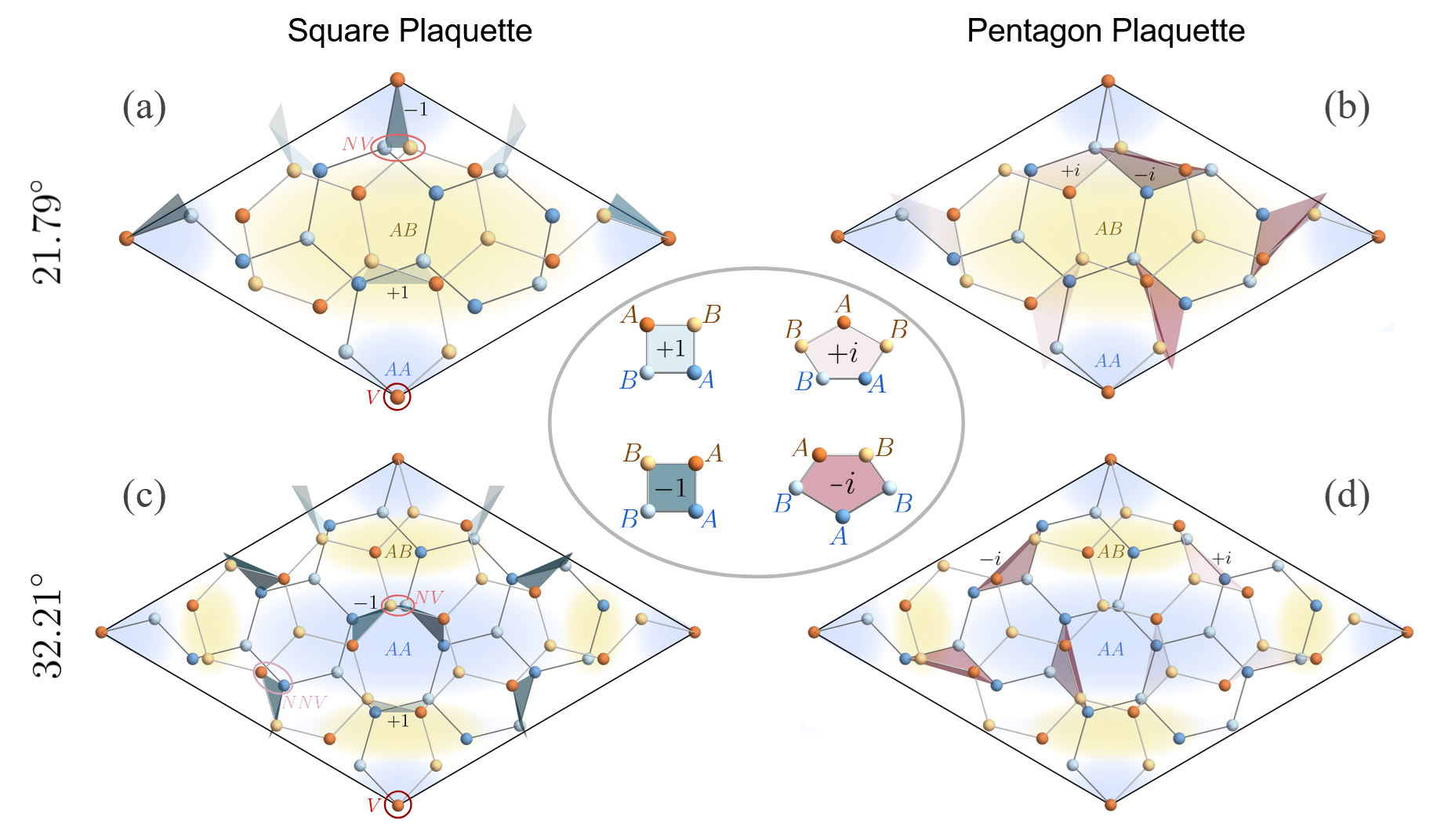}
        \caption{ Interlayer plaquette distribution in the moir\'e unit cell for two different twist angles. (a) Square and (b) pentagon interlayer plaquettes at twist angle $21.79^{\circ}$ around $C_{3z}$ invariant center and 
        (c) and (d) represent the same as (a) and (b) for the twist angle $32.21^{\circ}$. All the subfigures follow two different color schemes: one for background shades and other for plaquette types and plaquette eigenvalues. The background shades imply the stacking region: blue for AA region and yellow for AB region. For the plaquettes, the blue polygons depict square interlayer plaquettes, out of which light blue implies eigenvalue $+1$ and dark blue implies eigenvalue $-1$. Similarly, the red polygons depict pentagon plaquettes out of which light red pentagons imply $+i$ and dark red imply $-i$. The figure clearly shows all the $-1$ square plaquettes are in AA region, while $+1$ plaquettes are in AB region. The pentagon plaquettes are connected between AA and AB region. The plaquettes encircled in the middle illustrate the actual configuration of different plaquettes with their corresponding eigenvalues.}
    \label{fig:C3z_plaquette}
\end{figure*}
%
We next focus on the ground state phase diagram of moir\'e superlattices considering two different axes of twist passing through different centers. For an AA stacked honeycomb lattice, these centers correspond to: a) $C_{3z}$ invariant point, which coincides with lattice site and b) $C_{6z}$ invariant point positioned at the center of hexagon plaquette. In both cases, we employ the aforementioned Monte Carlo simulations to determine the minimum energy flux configuration.

\subsection{Ground state for twist around \texorpdfstring{$C_{3z}$}{C3z} invariant center}

\begin{figure*}[!t]
    \centering
    \includegraphics[width=0.98\textwidth]{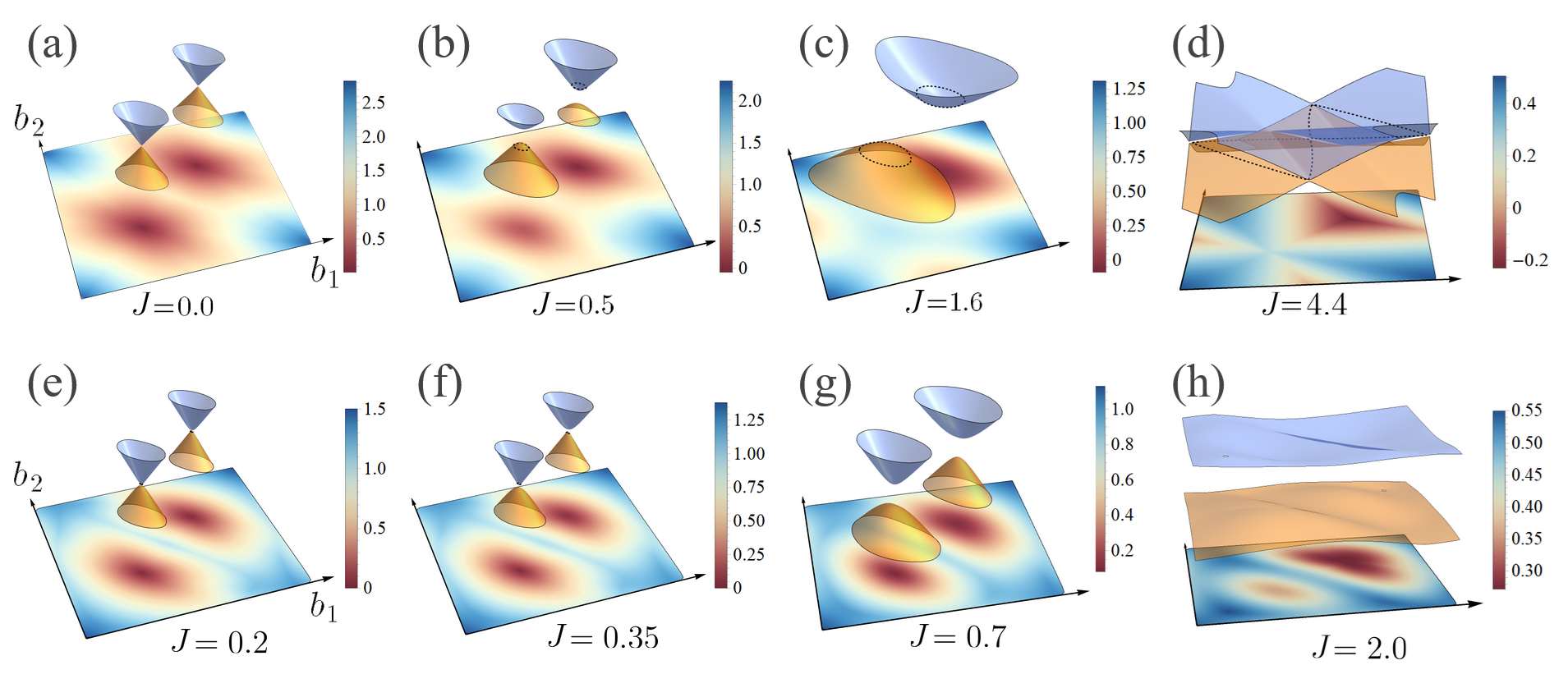}
        \caption{First row of the figure illustrates how the band spectrum changes with increasing interlayer strength J at twist angle $21.79^{\circ}$. Starting from the (a) gapless situation, (b)-(c) an indirect gap appears in the middle of the spectrum before the gapless state is restored again with increasing the coupling strength. Second row represents the same evolution at twist angle $32.21^{\circ}$. Here the spectrum transits from (e) gapless to (f) indirectly gapped configuration to a (g)-(h) fully gapped one as the interlayer strength is enhanced. The color plot at the base implies the energy value of conduction band over the Brillouin Zone. The black dashed line in the spectrum marks the zero energy value where the bands intersect it.
        }
    \label{fig:twist_energy.png}
\end{figure*}
We select a representative axis of twist passing through A sublattice sites in both the layers and focus only on two twist angles $\theta_{2,1} = 21.79^{\circ}$ and $\theta_{3,1} = 32.21^{\circ}$ around this axis. Twisting introduces both AA and AB stacking regions within the smallest moir\'e unit cell. These regions are distinguished by background shading, blue for AA stacking and yellow for AB stacking as shown in Fig.~\ref{fig:C3z_plaquette}. 
Additionally, we characterize four distinct plaquettes within the unit cell for each of these two twist angles. We use four different colors to represent these plaquettes with different eigenvalues as also encircled in the middle of Fig.~\ref{fig:C3z_plaquette}.

For the twist angle $21.79^{\circ}$, the unit cell contains one direct vertical (V) interlayer bond where the top atom aligns exactly above the bottom atom, and nine nearest-vertical bonds (NV) exhibiting slight lateral displacements. Representative example of each bond type is highlighted in Fig.~\ref{fig:C3z_plaquette}a by red circles. A detailed illustration and comprehensive Hamiltonian construction of this model is provided in  Appendix~\ref{Appendix_C}. The unit cell hosts six square\footnote{Note, the square is not a perfect square after twisting, rather a four-sided polygon. We kept the name to keep connection with previous section as shape has no impact in flux value.} plaquettes and six pentagon plaquettes as shown separately in Fig.~\ref{fig:C3z_plaquette}a and Fig.~\ref{fig:C3z_plaquette}b, respectively. For the minimum energy configuration, all three square interlayer plaquettes located within the blue shaded AA region take individually $-1$ value (dark blue square). This resembles the bilayer AA stacked case where all the interlayer plaquettes are in the AA regimes and take $-1$ value in the ground state. The rest three square plaquettes in the yellow shaded AB region carry $+1$ flux value (light blue square). Likewise, the six pentagon plaquettes connect AA and AB regions (blue to yellow) as evident from Fig.~\ref{fig:C3z_plaquette}b. Among them, each of three plaquettes in the left half of the unit cell carries flux value of $+i$ (light red pentagon), while their layer-exchanged reflection-symmetric counterparts on the right half exhibit $-i$ flux value (dark red pentagon).
\begin{figure*}[!t]
    \centering
    \includegraphics[width=0.99\textwidth]{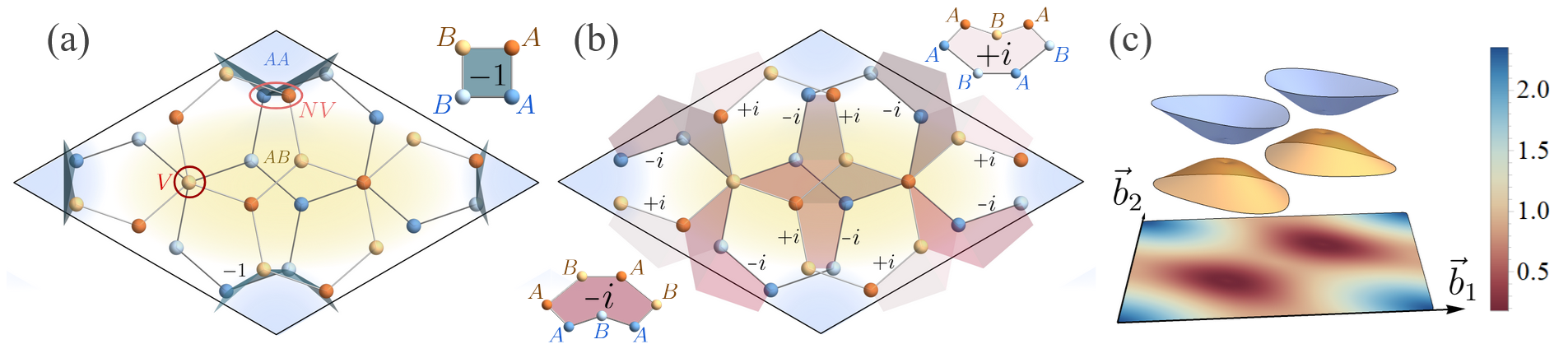}
        \caption{The unit cells show the plaquette configuration of (a) square (Blue) and (b) pentagon (Red) interlayer plaquettes at twist angle $21.79^{\circ}$ around $C_{6z}$ invariant point. Similar to Fig.~\ref{fig:C3z_plaquette}, the blue region in the unit cell is AA region and yellow region is AB/BA region. (c) The bulk-band spectrum shows a finite gap which will remain for any general $J$ value.
        }
    \label{fig:C6z}
\end{figure*}
A similar pattern of the interlayer plaquettes and their corresponding eigenvalues are also observed for the twist angle $32.21^{\circ}$. 
In this case, the moir\'e unit cell features three types of allowed interlayer bonds: one strictly vertical (V), nine nearest vertical (NV) and six next-nearest vertical (NNV) with progressively larger lateral displacements as illustrated in Fig.~\ref{fig:C3z_plaquette}c.
As such nine square plaquettes come up within the moir\'e unit cell as shown in Fig.~\ref{fig:C3z_plaquette}c.  Six of them located in the AA regime carry flux values $-1$ and the remaining three residing in the AB stacking regime have flux values $+1$. Fig.~\ref{fig:C3z_plaquette}d demonstrates pattern for the pentagon plaquettes. Out of six pentagons, each of the three pentagons carries flux value $-i$, and the rest three carry $+i$ value for each.

\begin{figure*}[!t]
    \centering
    \includegraphics[width=1.0\textwidth]{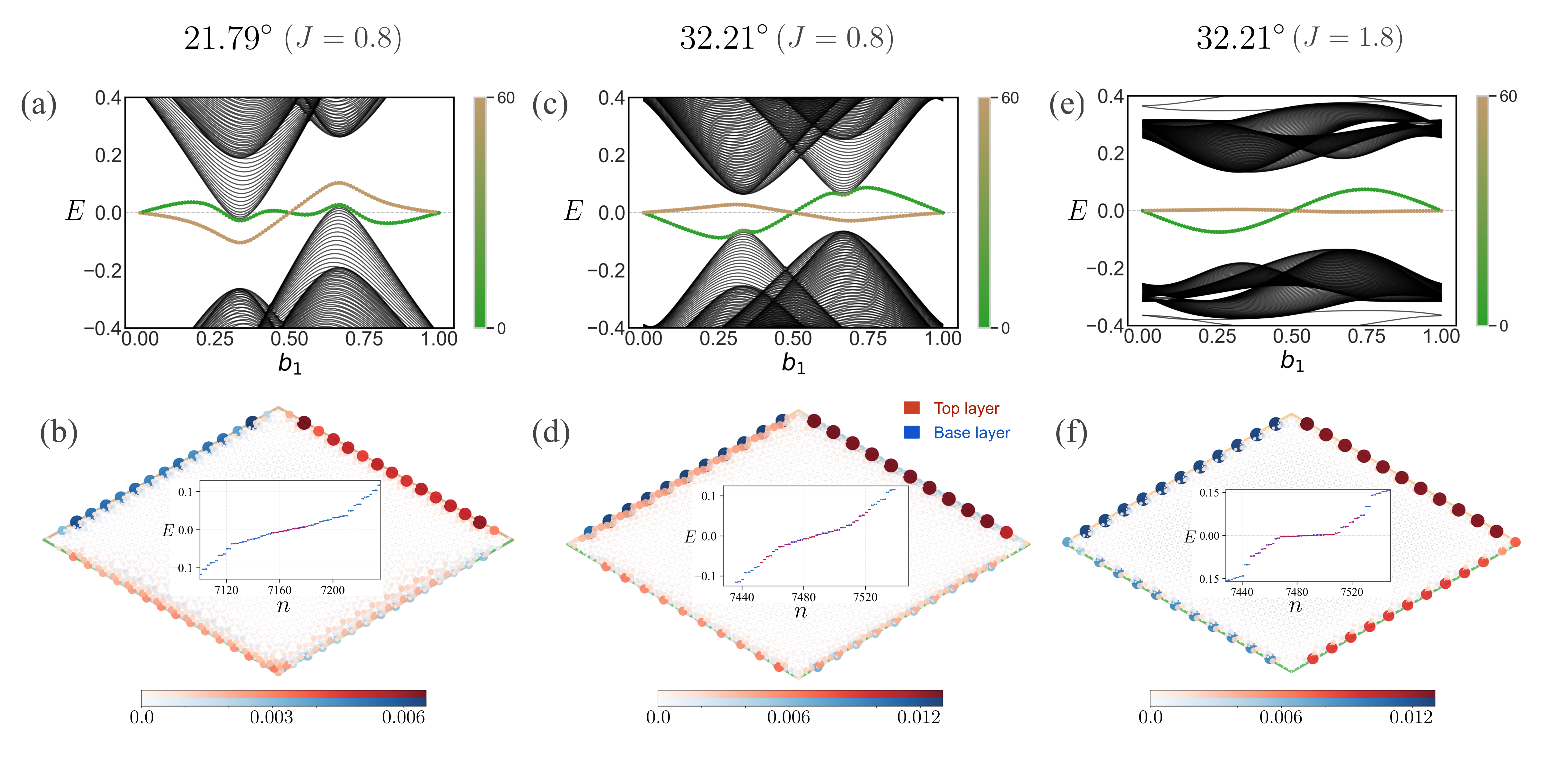}
        \caption{(a) Edge spectrum for a $100 \times 60$ lattice finite along second direction exhibiting two counter-propagating edge modes from opposite edges. (b) The spatial distribution of wave functions for energy values near the Fermi level (red dots in the energy $E$ vs eigenvalue index $n$ plot shown in the inset) for a $16 \times 16$ unit cell under open boundary condition (OBC). Both the figure are at $21.79^{\circ}$ twist angle under $C_{3z}$ rotation for $J = 0.8$. At $32.21^{\circ}$ twist angle, corresponding edge spectrum and wave function distribution for $12 \times 12$ unit cell are represented for (c-d) $J = 0.8$ and (e-f) $J=1.8$. In (a,c,e) the colorbar indicates the indexing of the unit cell along $\vec{a}_2$. In (b,d,f) the blue and red circles denote the wavefunction amplitude of base and top layers respectively.
        }
    \label{fig:C3z_Edge}
\end{figure*}
Next we analyze the energy spectrum for both twist angles discussed above. At $21.79^{\circ}$, the unit cell has 7 sites per layer, resulting in a $56 \times 56$ momentum space Hamiltonian due to two layers, two sublattices and two flavors of Majorana fermion at each site. Similarly, $32.21^{\circ}$ twist angle yields $104 \times 104$ momentum space Hamiltonian originating from 13 sites per layer. The corresponding energy spectra of the lowest conduction and highest valence bands for these twist angles are shown in first and second row of Fig.~\ref{fig:twist_energy.png},  respectively. At $J = 0$, we recover the gapless Dirac spectrum as shown in Fig.~\ref{fig:twist_energy.png}a. However, at $21.79^{\circ}$ and with a finite $J$, we obtain gapped Dirac nodes (Fig.~\ref{fig:twist_energy.png}b) due to breaking of sublattice symmetry. As we increase $J$, the valence and conduction bands shift in momentum, resulting in an indirect {\it negative} gapped spectrum (see Fig.~\ref{fig:twist_energy.png}c). If we increase $J$ further, the bandgap gradually decreases and closes at $J = 4.4$, leading to gapless spectrum as shown in Fig.~\ref{fig:twist_energy.png}d. In contrast, for the twist angle $32.21^{\circ}$, the band gap evolves differently (Fig.~\ref{fig:twist_energy.png}e-h). For $J < 0.42$, an indirect band gap is present (Fig.~\ref{fig:twist_energy.png}e-f), where both the valence and conduction bands cross Fermi energy. However, for $J>0.42$, we obtain a fully gapped insulating phase as depicted in Fig.~\ref{fig:twist_energy.png}g-h. 

\subsection{Ground state for twist around \texorpdfstring{$C_{6z}$}{C6z} invariant center}

The flux patterns discussed in the preceding section provide insights into how the plaquette configurations of the ground state in the AA and AB stacking regions change under twist around the axis through $C_{3z}$ invariant point. Does this scenario hold if we twist around any axis through different symmetry-invariant point of the honeycomb lattice? To address this, we next turn to investigate the plaquette configurations of the ground state for $21.79^{\circ}$ twist around the {\it center} of the hexagon plaquette of the AA stacked case. As before, the unit cell also features two distinct types of interlayer bonds: two V and six NV. The representative bond of each type is shown in Fig.~\ref{fig:C6z}a. The positions of the plaquettes are redistributed throughout the unit cell. Fig.~\ref{fig:C6z} shows that all six square interlayer plaquettes reside in AA region as opposed to the $C_{3z}$ case, and carry same flux value $-1$ in the ground state configuration as before. Additionally, the shortest odd-length plaquette turns out to be heptagon in contrast to the $C_{3z}$ case. Fig.~\ref{fig:C6z}b depicts that there are twelve heptagons within the unit cell, spanning from AA to AB regions. Each of the two V bonds in the AB region is surrounded by six heptagons: three of them acquire $+i$ value (light red polygon) and the other three individually acquire $-i$ value (dark red polygon) in a periodic pattern. Six other reflection-symmetric heptagons in another half of the unit cell will carry conjugate values of the first six. We also plot the energy spectrum of the lowest conduction and highest valence bands in Fig.~\ref{fig:C6z}c for this specific flux configuration. For any finite $J$ value, the spectrum remains fully gapped, suggesting an insulating bulk character of the spin liquid phase.

\subsection{Edge states}
\par
\begin{figure}
    \centering
    \includegraphics[width=1.0\columnwidth]{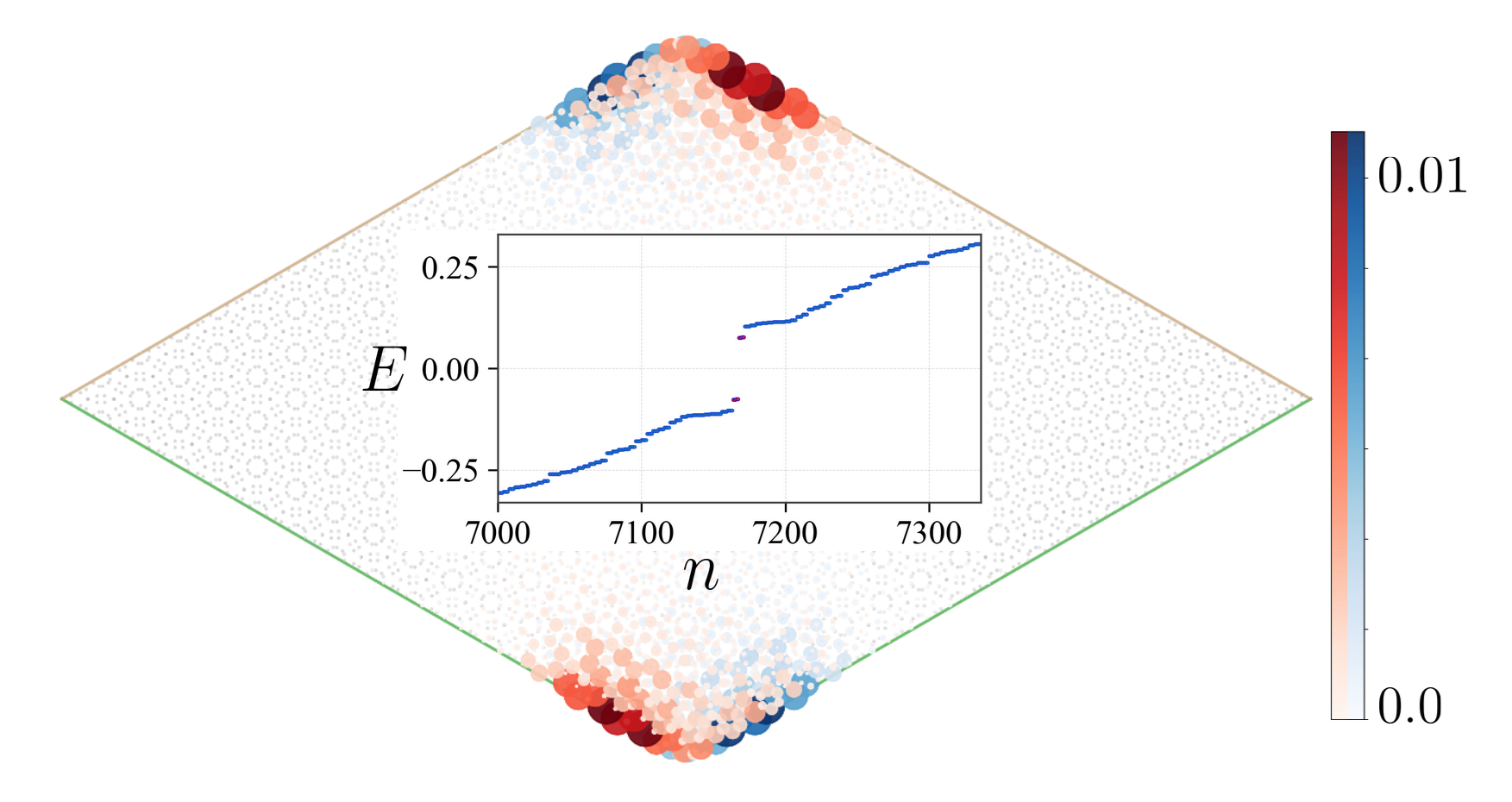}
        \caption{The corner states associated with the distinct red shaded closest energy eigenvalues at $21.79^{\circ}$ twist angle under $C_{6z}$ invariant rotation for $16 \times 16$ lattice with OBC.
        } 
    \label{fig:C6z_Edge}
\end{figure}

 Having discussed the ground states of spin-liquids in Kitaev bilayers and moir\'e superlattices, we now check if there exists any non-trivial edges particularly in moir\'e superlattices due to the breaking of time-reversal symmetry. For convenience, we focus on twist around $C_{3z}$ for the two twist angles $21.79^{\circ}$ and $32.21^{\circ}$. We consider a $100 \times 60$ lattice with a finite cut along the basis vector  $a_2$, keeping periodicity along  $a_1$. The bulk and boundary spectrum in Fig.~\ref{fig:C3z_Edge}a shows the existence of two counter-propagating modes from the two opposite surfaces for a representative value of the interlayer coupling ($J=0.8$) within the regime, where bulk spectrum has an indirect gap. Out of these two modes, one connects delocalized bulk conduction and valence bands which we refer as {\it connected} boundary mode. The other mode turns out to float between conduction and valence bands which we refer as {\it disconnected} mode. The existence of such a disconnected boundary mode is also observed in other models \cite{zhang2024probing,FragilityAltland2024,Cook2022} which are robust against backscattering. Since the nature of these boundary modes is different from the standard boundary modes, the topological characterization of these modes is beyond the scope of the current study. We now discuss the spatial distribution of the wavefunctions of these boundary modes for a 2D bilayer lattice with open boundaries as shown in Fig.~\ref{fig:C3z_Edge} (b,d,f). Interestingly, the disconnected boundary mode near the zero energy is strongly localized along its corresponding edges of the finite lattice as compared to the connected boundary mode as shown in Fig.~\ref{fig:C3z_Edge}b. For the twist angle $32.21^{\circ}$ with the same twist center and interlayer coupling strength, the spectrum of the boundary modes and the spatial distribution of these modes near zero energy show similar behavior as evident from Fig.~\ref{fig:C3z_Edge}c. However, both the boundary modes are found to be disconnected from the valence and conduction bands as we enhance interlayer strength to $J = 1.8$. This is clearly visible in Fig.~\ref{fig:C3z_Edge}e. Accordingly, the spatial distribution of the wavefunctions starts to grow on the other edges of the lattice (see Fig.~\ref{fig:C3z_Edge}f). In contrast to the $C_{3z}$ case, twist under   $C_{6z}$ center highlights presence of corner modes associated with two separated degenerate eigenvalues near Fermi energy as shown in Fig.~\ref{fig:C6z_Edge}, indicating possible emergence of higher-order topology. However, this requires a comprehensive study, which we leave for future.   

\section{\label{Section_4} Conclusions}
We have proposed an exactly solvable spin-liquid in Kitaev bilayer and moir\`e superlattices. Using variational approach and Monte Carlo simulations, we have found that such a bilayer model can host various vison crystal phases with distinct interlayer and intralayer flux configurations in the presence of an out-of-plane magnetic field. Once twist is introduced, the ground state turns out to exhibit both even and odd-sided interlayer plaquettes with fluxes $\pm 1$ and $\pm i$, respectively. Remarkably, this interlayer flux configurations follow a generalized local stacking‐dependent pattern in the AA and AB regions, regardless of twist angle and twist center. We have further shown the emergence of gapless and gapped spin-liquid phases depending on the angle of twist and center of twist. Finally, we have shown the presence of an interesting floating edge mode in the twisted models in contrast to the standard edge modes. However, the reason for the emergence of such floating mode is not obvious and the topological characterization of these modes requires detailed understanding. Other potentially interesting future directions include investigating the fate of the spin liquid phase under generic non-Kitaev type spin interaction as well as extending this model to any arbitrary twist angle in different twisted bilayer systems.

\section{Acknowledgments}
ID would like to thank Saptarshi Mandal for useful discussions. ID also acknowledges VIRGO and KALINGA clusters where most of the numerical calculations were performed. OE acknowledges support from NSF Award No. DMR-2234352.

\section{Data Availability Statement}
The datasets underlying the plots that support this study’s conclusions are available upon reasonable request from the authors.

\appendix

\section{\label{Appendix_A} Majorana fermionization of the bilayer Hamiltonian}
\begin{figure*}[!t]
    \centering
    \includegraphics[width=0.93\textwidth]{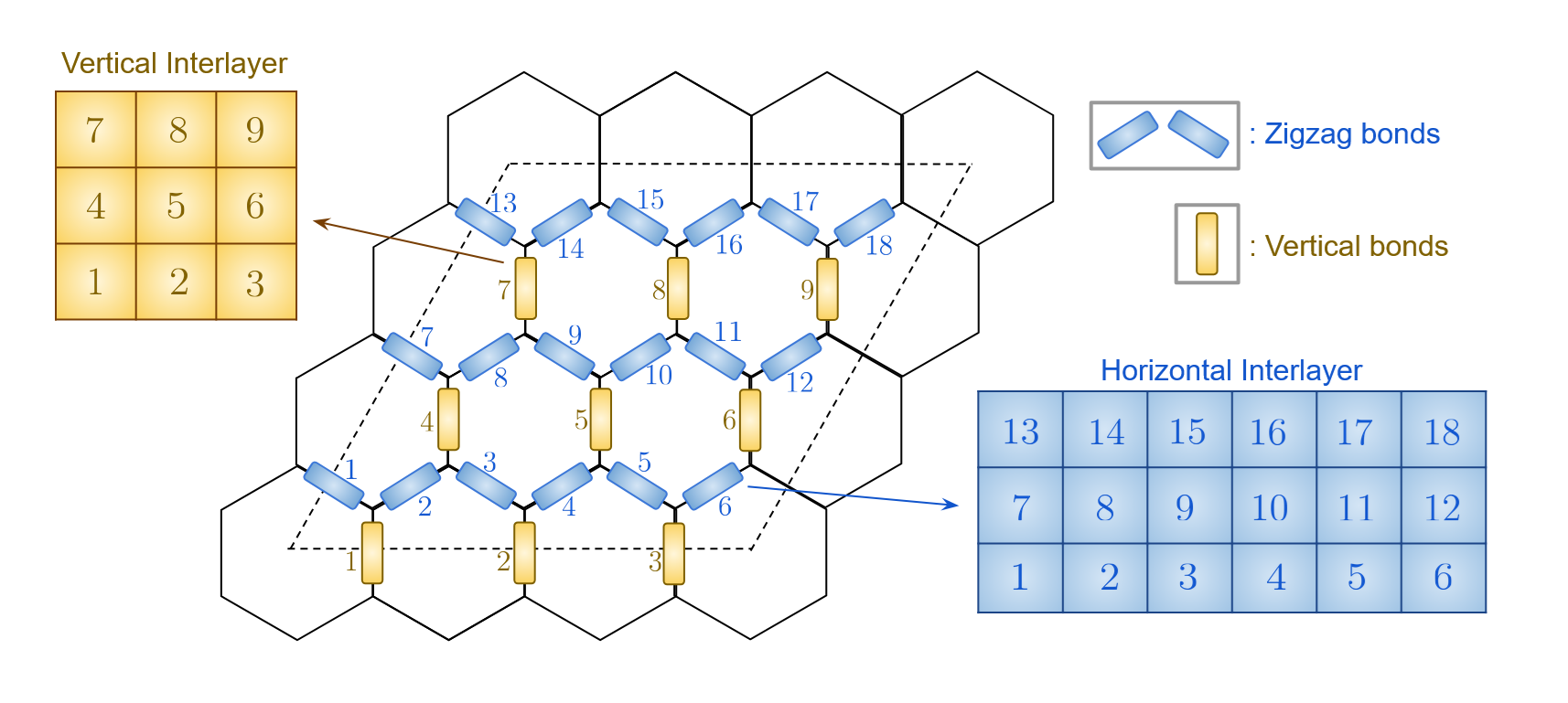}
        \caption{Pictorial characterization scheme of interlayer plaquettes within a representative $3 \times 3$ unit cell marked by dashed line.}
    \label{fig:Interlayer_Scheme}
\end{figure*}
The objective of this Appendix is to derive Eq.~\ref{eq:Majorana_Gamma_Model} in the main text. We begin with the Majorana representation of Gamma matrices as $\Gamma_j^\gamma = i\,b_j^\gamma\,c_j$, where $b^{\gamma}$ and $c$ are Majorana operators. They are hermitian in nature and satisfy standard anticommutation relation $\{c_i, c_j\}_{i \ne j} = 0$ and $c_i^2 =1$ with analogous relation for $b$'s. Here all the Majoranas can be treated in the same basis regardless of the site index. Using these standard commutation relations and definition of bilayer operators $\Gamma^{\gamma \gamma'} = \frac{i}{2}[\Gamma^{\gamma}, \Gamma^{\gamma'}]$, the Majorana representation of the bilinear operators are found to be
$\Gamma_j^{\gamma\gamma'} = i\,b_j^\gamma\,b_j^{\gamma'}$. Then the total Hamiltonian takes the form
\begin{widetext}
\begin{align}
 \mathcal{H} &=  \sum_{\langle jk\rangle_{\gamma},\nu} K_\gamma\; \{(i\,b_{\nu j}^\gamma\,c_{\nu j})\,(i\,b_{\nu k}^\gamma\,c_{\nu k})\; +\;(i\,b_{\nu j}^\gamma\,b_{\nu j}^5)\,(i\,b_{\nu k}^\gamma\,b_{\nu k}^5)\} + J\;\sum_{j} \{(i\,b_{1 j}^4\,c_{1 j})\,(i\,b_{2 j}^4\,c_{2 j})\; +\;(i\,b_{1 j}^4\,b_{1 j}^5)\,(i\,b_{2 j}^4\,b_{2 j}^5)\}\nonumber\\
 &= -\sum_{\langle jk\rangle_{\gamma},\nu} K_\gamma\; \{(i\,b_{\nu j}^\gamma\,b_{\nu k}^{\gamma})\,(i\,c_{\nu j}\,c_{\nu k})\; +\;(i\,b_{\nu j}^\gamma\,b_{\nu k}^{\gamma})\,(i\,b_{\nu j}^5\,b_{\nu k}^5)\} - J\;\sum_{j} \{(i\,b_{1 j}^4\,b_{2 j}^4)\,(i\,c_{1 j}\,c_{2 j})\; +\;(i\,b_{1 j}^4\,b_{2 j}^4)\,(ib_{1 j}^5\,\,b_{2 j}^5)\}
\label{Supply_Eq:Majorana_Gamma_Model}.
\end{align}

By relabeling the Majorana $b_j^5 \rightarrow c_j^x$ and $c_j \rightarrow c_j^y$, the Hamiltonian can be recasted as,
\begin{align}
H  = -\; \sum_{\nu, \langle jk \rangle} iK_{\gamma}\;u^{\gamma}_{\nu,jk}\;(c^{x}_{\nu j}\,c^{x}_{\nu k}\; +\;c^{y}_{\nu j}\,c^{y}_{\nu k}) 
- J\;\sum_{j}i\;u^{4}_{j}\;(c^{x}_{1j}\,c^{x}_{2j}\; +\;c^{y}_{1j}\,c^{y}_{2j}).
\label{eq:Majorana_Ham_Model_appendix}
\end{align}
\end{widetext}
Here, $u^{\gamma}_{\nu,jk} = i b_{\nu j}^{\gamma} b_{\nu k}^{\gamma}$ for $\gamma \in (1,2,3)$ and $u^{4}_{j} = i b_{1j}^{4} b_{2j}^{4}$ both acts as $\mathbb{Z}_2$ gauge fields in intra and interlayers respectively with the eigenvalues $\pm 1$. Also in this representation, the external field is expressed as
\eq{
H_h = h_z \sum_{\nu,j} \Gamma_{\nu j}^5 =  h_z \sum_{\nu,j} ib_{\nu j}^5 c_{\nu j} \equiv h_z \sum_{\nu,j} i c_{\nu j}^x c_{\nu j}^y.
\label{eq:Zeeman_Term_appendix}
}
Equations \ref{eq:Majorana_Ham_Model_appendix} and \ref{eq:Zeeman_Term_appendix} are quadratic in Majorana operators, hence can be solved exactly. As the fermionic Hilbert space per site is $2^3 = 8$ dimensional, twice the size of the Gamma matrices, we restrict the physical states by imposing the local constraint $D_{\nu j} \ket{\psi} = +1 \ket{\psi}$, where
\begin{align}
D_{\nu j} = \Gamma^1_{\nu j} \Gamma^2_{\nu j} \Gamma^3_{\nu j} \Gamma^4_{\nu j} \Gamma^5_{\nu j} = i b^1_{\nu j} b^2_{\nu j} b^3_{\nu j} b^4_{\nu j} c^x_{\nu j}c_{\nu j}^y.    
\end{align}

\begin{figure*}[!t]
    \centering
    \includegraphics[width=0.8\textwidth]{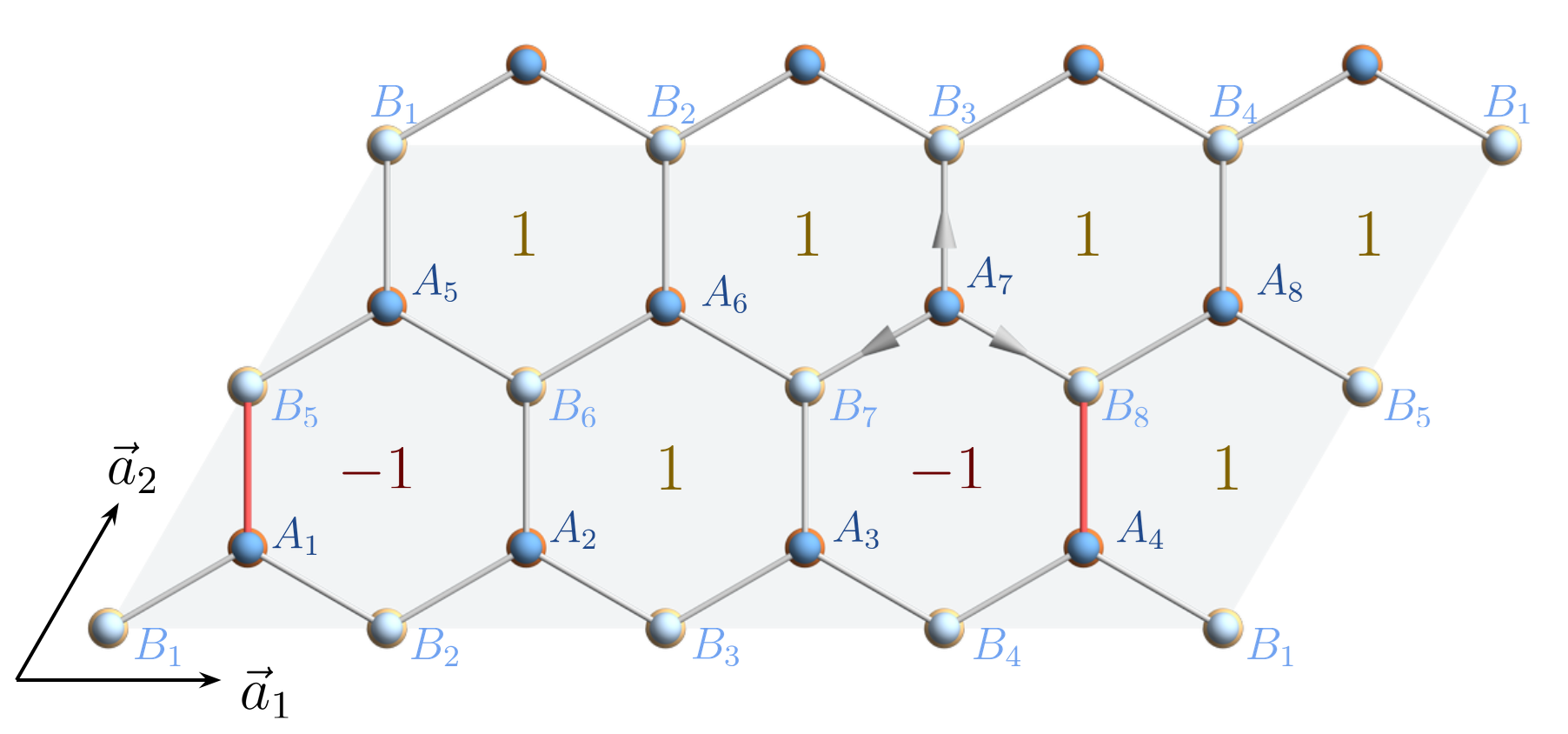}
        \caption{The unit cell of $1/4 - \pi$ flux phases. There exists 8 sites in each layer with two sublattices A and B for each. Generally throughout the unit call  $u_{\nu, AB} = +1$, except the red bonds where $u_{\nu, AB} = -1$. Here, $\vec{a}_1 = (4\sqrt{3},0)a$ and $\vec{a}_2 = (\sqrt{3},3)a$ are the lattice vectors of the unit cell in real space. }
    \label{fig:one_fourth_flux}
\end{figure*}

\section{\label{Appendix_A1} Characterization scheme for interlayer plaquettes}
The distribution of interlayer plaquette eigenvalues within the unit cell is difficult to visualize. To represent them conveniently, we introduce a simplified scheme as used in Fig.~\ref{fig:Different_plaquette} of the main text. Fig.~\ref{fig:Interlayer_Scheme} illustrates the scheme which involves identifying horizontal bonds as zigzag bonds (shown in blue), keeping vertical bonds as it is (shown in yellow). Corresponding eigenvalues associated with the interlayer plaquettes in the third direction are portrayed through horizontal and vertical interlayer matrices respectively. Note that, we separate interlayer plaquettes as zigzag and vertical bonds solely for visual convenience. However to define interlayer flux phase we consider total number of interlayer plaquettes involving both zigzag and vertical bonds.


\section{\label{Appendix_B} Construction of quadratic Majorana Hamiltonian for different flux sectors}
The objective of this appendix is to construct Majorana Hamiltonian in momentum space, which was used to find excitation spectrum (right column of Fig.~\ref{fig:Different_plaquette}) for different flux phases presented in Fig.~\ref{fig:Different_plaquette} of the main text. We particularly focus on the 
$\frac{1}{4}$-$\pi$ flux phase and comment on the other phases. 
The quadratic Majorana tight-binding Hamiltonian in Eq.~\ref{eq:Majorana_Ham_Model_appendix} can be recasted as  
\eq{
\mathcal{H} = i \sum_{\substack{j,s_1,\nu,\alpha;\\k,s_2,\nu',\beta}} c_{j,s_1,\nu,\alpha}\;\; \mathcal{A}_{\substack{j,s_1,\nu,\alpha;\\k,s_2,\nu',\beta}} \;\;  c_{k,s_2,\nu',\beta},
}
where $j,k \in (1,2,...,N_1 N_2)$ label the sites for $N_1 \times N_2$ finite lattice, $s \in A,B$ denotes sublattices, $\alpha \in (x,y)$ represents two types of free Majorana fermions, $\nu \in 1,2$ distinguishes base and top layer respectively, and ${\mathcal A}$ is the  $8N_1N_2 \times 8N_1N_2$ matrix involving $ K_{\gamma}\;u^{\gamma}_{\nu,jk}, J,  u^{4}_{j}$ as the matrix elements. As seen in Fig.~\ref{fig:Different_plaquette} of the main text, the effective unit cell for different flux phases expands in the honeycomb lattice due to the symmetry breaking of the system except the $0-\pi$ flux phase (see Fig.~\ref{fig:Fig2} of the main text). Fig.~\ref{fig:one_fourth_flux} illustrates unit cell for the $1/4-\pi$ phase and we choose the gauge in the intralayer sectors as $u_{\nu, AB} = +1$ for all bonds except two red bonds, where the sign is reversed. This gives two $-1$ plaquettes out of eight plaquettes, resulting in one-fourth flux contribution. For interlayer links, we adopt $u_{A} = -1$ and $u_{B} = 1$, which effectively imposes $\pi$ flux on all square plaquettes.

Using this gauge choice, we solve the tight-binding Hamiltonian by writing  Majorana operators in Fourier basis as
\eq{
c_{j,s_1,\nu,\alpha} = \sqrt{\frac{2}{N}} \sum_{\mathbf{k} \in \frac{BZ}{2}} [ c_{\mathbf{k},s_1,\nu,\alpha} e^{i \mathbf{k}.\mathbf{x}_i} + c_{\mathbf{k},s_1,\nu,\alpha}^{\dag} e^{-i \mathbf{k}.\mathbf{x}_i}].\nonumber
}
While $c_j^{\dag} = c_j$ in real space, in momentum space they obey $c_{\mathbf{k}}^{\dag} = c_{-\mathbf{k}}$. The bond vectors for our model are given by $(\pm \frac{\sqrt{3}}{2},-\frac{1}{2})a,(0,1)a$, $a$ being the bond length considered as unity for all bonds in AA stacked lattice.
\begin{figure*}[!t]
    \centering
    \includegraphics[width=0.78\textwidth]{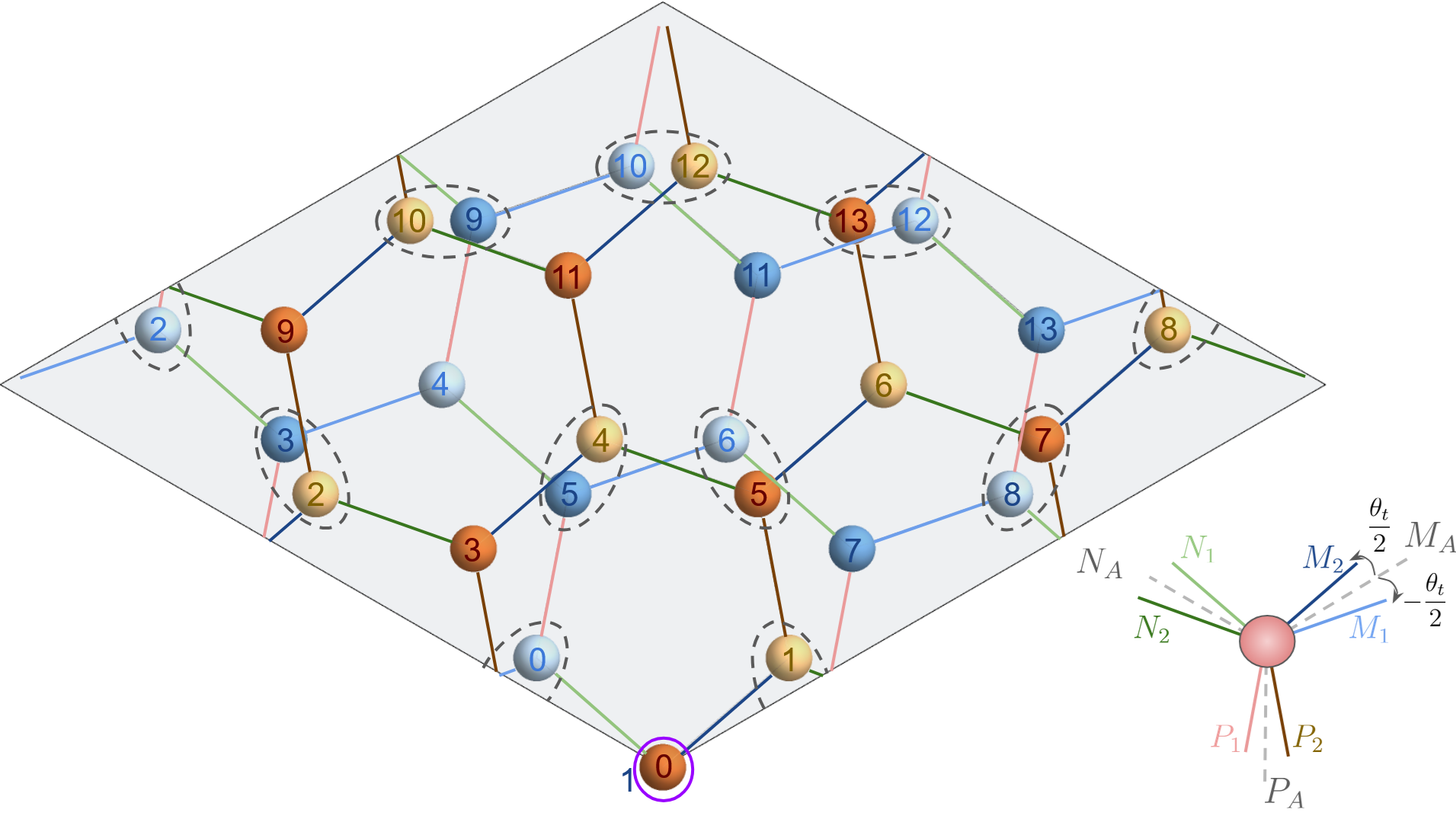}
        \caption{The unit cell of moir\'e superlattice with twist angle $\theta_t = 21.8^0$. The blue (yellow) layer represents the base (top) layer with a twist angle of $\mp\theta_t/2$. While the light intralayer bonds are for base layers ($M_1:$ Blue, $N_1:$ green, $P_1:$ Red), the dark bonds ($M_2, N_2, P_2$) are for top layers. The dashed lines creating the pair between base and top layers are nearest-vertical (NV) interlayer bonds along with the pink vertical bond (V) at the bottom. The inset shows the rotated intralayer bonds of base and top layers with respect to AA stacked intralayer bonds (dashed bonds named as $M_A$,$N_A$,$P_A$)  }
    \label{fig:indexed_moire_21}
\end{figure*}

Assuming $K_{\gamma} = K$, we define the relevant matrix elements in $\mathcal{A}$ in momentum space as follows:
\begin{widetext}
\begin{align}
M = 2 i K\, \text{exp}[ia(-\frac{\sqrt{3}}{2}k_x - \frac{1}{2}k_y)],\qquad N = 2 i K\, \text{exp}[ia(\frac{\sqrt{3}}{2}k_x - \frac{1}{2}k_y)],\qquad P =  2 i K\, \text{exp}[iak_y]
\end{align}
In the basis $(c_{A,1}^x, c_{A,1}^y, c_{B,1}^x, c_{B,1}^y, c_{A,2}^x, c_{A,2}^y, c_{B,2}^x, c_{B,2}^y, ...,c_{B,8}^x, c_{B,8}^y)^T$, the intralayer Hamiltonian matrix for the $4 \times 2$ unit cell shown in Fig.~\ref{fig:one_fourth_flux} is given by

\begin{align}
H_{\text{intra}}(\mathbf{k})  = 
\begin{bmatrix}
\begin{array}{cccccccccccccccc}
0 & M & 0 & N & 0 & 0 & 0 & 0 & 0 & -P & 0 & 0 & 0 & 0 & 0 & 0 \\  
0 & 0 & 0 & 0 & 0 & 0 & 0 & 0 & 0 & 0 & 0 & 0 & 0 & 0 & 0 & 0 \\
  &   &   & M &   & N &   &   &   &   &   & P  &   &   &   &  \\  
\vdots &  &  &  &  &  & \vdots & \vdots &  &  &  &  &  &  &  \vdots&  \\
 &  &  &  &  &  &  & P &  & N &  &  &  &  &  & M \\
0 & 0 & 0 & 0 & 0 & 0 & 0 & 0 & 0 & 0 & 0 & 0 & 0 & 0 & 0 & 0 \\
\end{array}
\end{bmatrix}
\otimes \mathbb{I}_2,   
\label{eq:1/4intra}
\end{align}

Similarly, the interlayer matrix is

\begin{align}
H_{\text{inter}}(\mathbf{k})  = 2iJ
\begin{bmatrix}
\begin{array}{cccccccccccccccc}
-1 &  &  &   &   &   &   &   &   &   &   &   &   &   &   &   \\  
  & 1 &   &   &   &   &   &   &   &  &  &   &   &   &   &   \\
 &  & -1 &   &   &   &   &   &   &   &   &   &   &   &   &   \\  
  &  &   & 1  &   &   &   &   &   &  &  &   &   &   &   &   \\
 &  &  &  & \ddots &  &  &  &  &  &  &  &  &  &  &  \\
\end{array}
\end{bmatrix}
\otimes \mathbb{I}_2
\label{eq:1/4inter}
\end{align}
\end{widetext}
$\mathbb{I}_2$ for both Eq.~\ref{eq:1/4intra} and \ref{eq:1/4inter} signifies two decoupled Majorana fermions. Then the full field-free Hamiltonian for the $1/4-\pi$ phase can be expressed as 
\begin{align}
H_{\frac{1}{4}-\pi}\bf(k)  = \left(\begin{matrix}H_{\text{intra}}& H_{\text{inter}}\cr 0& H_{\text{intra}} \end{matrix}\right) + \text{H.c.}   
\label{eq:1/4total}
\end{align}

The out-of-plane Zeeman term introduces coupling between the Majorana fermions. Under external field $h_z$,
\begin{align}
H_{\text{h}}(\mathbf{k})  =  \mathbb{I}_{32} \otimes
\begin{bmatrix}
\begin{array}{cc}
0 & 2ih_z  \\  
-2ih_z  & 0 \\
\end{array}
\end{bmatrix}
\label{eq:1/4h}
\end{align}
adds up with $H_{\frac{1}{4}-\pi}\bf(k)$ to give full Hamiltonian. $\mathbb{I}_{32}$ comes from 16 atomic sites considering both layers with two sublattices for each. Following this procedure, the Hamiltonian of other flux phases can also be constructed. However, the unit cell and corresponding matrix dimension of the momentum space Hamiltonian will differ accordingly.



\section{\label{Appendix_C} Twisted bilayer Hamiltonian at 21.79 degree}

This appendix provides a detailed construction scheme of moir\'e Hamiltonian as used in Sec.~\ref{sec:moire} for twist angle  $\theta_t = 21.79^{\circ}$ for twist around $C_{3z}$ invariant point. Fig.~\ref{fig:indexed_moire_21} illustrates that each layer of moir\'e unit cell at $21.79^{\circ}$ comprises seven sites partitioned equally between two sublattices, A and B, yielding a total of 14 sites per layer for each Majorana fermion. To visualize this, we color A and B sublattices of base layer dark and light blue respectively and corresponding sublattices of top layer as red and yellow. In the base layer, the intralayer bonds ($M_1, N_1, P_1$ in inset) are depicted in light tones, whereas those in the top layer ($M_2, N_2, P_2$ in the inset) are shown in dark tones. The AA stacked intralayer bond vectors (dashed bonds) shown in this inset are given by
\begin{align}
 M_{A} &= (\frac{\sqrt{3}}{2}, \frac{1}{2})a,\nonumber\\
 N_{A} &= (-\frac{\sqrt{3}}{2}, \frac{1}{2})a,\nonumber\\
 P_{A} &= (0, -1)a.
\end{align}
In the moir\'e configuration, these intralayer bonds transform as $M_{1,2} = \mathcal{R}(\mp \frac{\theta_t}{2}) M_A  \equiv \mathcal{R}_{-,+} M_A$ with similar transformations applied to $N_A$ and $P_A$, where $\mathcal{R}$ represents standard two-dimensional rotation matrix and $\mp$ refers to rotation along clockwise and anticlockwise direction respectively. Note that for all the bond vectors we present only the lateral coordinates ($x$,$y$) and omit the $z$-coordinate as it does not change under rotation.

\par Now regarding the interlayer coupling, here we categorize three distinct types of bonds depending on their bondlength $r$: direct vertical V $(r=a)$, nearest vertical NV $(1.07a)$ and next-nearest vertical NNV $(1.2a)$. In Fig.~\ref{fig:indexed_moire_21}, only one V bond $1_b \rightarrow 0_t$ (b: base site, t: top site) exists, encircled in pink. However, there are nine NV bonds as indicated by dashed circles between sites of base and top layers (e.g., $5_b \rightarrow 4_t$, $2_b \rightarrow 1_t$, etc). Additionally, there are several interlayer bonds which qualify as NNV bonds based on nearest-neighbor criterion. For example, $7_b \rightarrow 5_t$ is an NNV bond, but $5_t$ is already connected to $6_b$ as NV bond. We therefore impose a constraint on number of interlayer bonds such that each site can have at most one interlayer bond. Specifically, we introduce a cutoff length  $1.07<r_c<1.2a$ such that we can eliminate all NNV bonds in the current setting as discussed in Sec.~\ref{sec:moire}A of the main text . 
This preserves the exact solvability of the Hamiltonian. With this, we determine the ground state of the real space Hamiltonian using Monte Carlo simulation. It turns out the bond operator corresponding to V reverses sign for the ground state. As three square plaquettes out of six are connected with this bond (see Fig.~\ref{fig:C3z_plaquette}a) they take $-1$ value.

\par Now to construct the momentum space Hamiltonian of this minimum energy configuration, the translation vectors of the unit cell are taken to be $\vec{a}_1 = \mathcal{R}_+ (\frac{5 \sqrt3}{2}, \frac{3}{2})$ and $\vec{a}_2 = \mathcal{R}_- (-\frac{5 \sqrt3}{2}, \frac{3}{2}) \equiv (-a_{1x}, a_{1y})$. Accordingly, the reciprocal lattice vectors are found to be $\pi(\pm a_{1x}^{-1}, a_{1y}^{-1})$ which defines the Brillouin zone (BZ). Then we formulate the complete $56 \times 56$ momentum-space Hamiltonian within the BZ in terms of intra and interlayer as follows
\begin{align}
H_{\theta\rightarrow 21.79^{\circ}}\bf(k)  = \left(\begin{matrix}H_{\text{base}}& H_{\text{inter}}\cr H^*_{\text{inter}}&H_{\text{top}} \end{matrix}\right) \otimes  \mathbb{I}_2 .  
\label{eq:21_Ham_k_Suppli}
\end{align}
Similar to Eq.~\ref{eq:1/4intra} and \ref{eq:1/4inter}, the identity matrix refers to two Majorana flavors. The matrix elements of $28 \times 28$ intralayer block corresponding to $M_{1,2}$ bonds are
\begin{align}
M_{1,2}:\; 2 i K\, \text{exp}[i\vec{M}_{1,2}.\vec{k}].
\end{align}
Note that similar relations also hold for $N_{1,2}$, $P_{1,2}$. For example, the matrix element connecting sites $5_t$ and $1_t$  linked by $P_2$ bond is given by $H_{\text{top}}^{5,1} = 2iKu_{\langle 5,1 \rangle}\, e^{i (\vec{P}_{2}.\mathbf{k})}$ where $u_{\langle 5,1 \rangle}$ is the bond operator. Similarly for interlayer block Hamiltonian, the matrix element between site $5_b$ and site $4_t$ is described by $H_{\text{inter}}^{5,4} = 2iJu_{\langle 5_b,6_t \rangle}\, e^{i (\vec{r}_{4t}- \vec{r}_{5b}).\mathbf{k}}$, where $\vec{r}_{4t}$ and $\vec{r}_{5b}$ are the position vectors of the respective sites. Likewise, we can construct Hamiltonians for different twist angles and twist axes following this scheme.


\bibliography{references}

\end{document}